\documentclass[pdflatex,sn-nature]{sn-jnl}

\setcitestyle{super} 

\usepackage{graphicx}%
\usepackage{multirow}%
\usepackage{amsmath,amssymb,amsfonts}%
\usepackage{amsthm}%
\usepackage{mathrsfs}%
\usepackage[title]{appendix}%
\usepackage{xcolor}%
\usepackage{textcomp}%
\usepackage{manyfoot}%
\usepackage{booktabs}%
\usepackage{algorithm}%
\usepackage{algorithmicx}%
\usepackage{algpseudocode}%
\usepackage{listings}%
\usepackage{mathtools}
\DeclarePairedDelimiter\ceil{\lceil}{\rceil}

\usepackage{float}
\usepackage{color,soul}
\usepackage[version=4]{mhchem} 

\usepackage{xr}
\makeatletter
\newcommand*{\addFileDependency}[1]{
  \typeout{(#1)}
  \@addtofilelist{#1}
  \IfFileExists{#1}{}{\typeout{No file #1.}}
}
\makeatother

\newcommand*{\myexternaldocument}[1]{%
    \externaldocument{#1}%
    \addFileDependency{#1.tex}%
    \addFileDependency{#1.aux}%
}

\myexternaldocument{supp_info} 

\theoremstyle{thmstyleone}%
\theoremstyle{thmstyletwo}%

\theoremstyle{thmstylethree}%

\begin{document}

\title[Article Title]{Limitations of Quantum Hardware for Molecular Energy Estimation Using VQE}


\author*[1]{\fnm{Abel} \sur{Carreras}}\email{abelcarreras83@gmail.com}

\author[1,2,3]{\fnm{Rom\'an} \sur{Or\'us}}\email{roman.orus@dipc.org}

\author[1,3]{\fnm{David} \sur{Casanova}}\email{david.casanova@dipc.org}

\affil[1]{\orgname{Donostia International Physics Center (DIPC)}, \orgaddress{\city{Donostia}, \postcode{20018}, \state{Euskadi}, \country{Spain}}}

\affil[2]{\orgname{Multiverse Computing}, \orgaddress{\city{Donostia}, \postcode{20014}, \state{Euskadi}, \country{Spain}}}

\affil[3]{\orgname{IKERBASQUE, Basque Foundation for Science}, \orgaddress{\city{Donostia}, \postcode{48009}, \state{Euskadi}, \country{Spain}}}


\abstract{Variational quantum eigensolvers (VQEs) are among the most promising quantum algorithms for solving electronic structure problems in quantum chemistry, particularly during the Noisy Intermediate-Scale Quantum (NISQ) era. In this study, we investigate the capabilities and limitations of VQE algorithms implemented on current quantum hardware for determining molecular ground-state energies, focusing on the adaptive derivative-assembled pseudo-Trotter ansatz VQE (ADAPT-VQE). To address the significant computational challenges posed by molecular Hamiltonians, we explore various strategies to simplify the Hamiltonian, optimize the ansatz, and improve classical parameter optimization through modifications of the COBYLA optimizer. These enhancements are integrated into a tailored quantum computing implementation designed to minimize the circuit depth and computational cost.  
Using benzene as a benchmark system, we demonstrate the application of these optimizations on an \texttt{IBM} quantum computer. Despite these improvements, our results highlight the limitations imposed by current quantum hardware, particularly the impact of quantum noise on state preparation and energy measurement. The noise levels in today's devices prevent meaningful evaluations of molecular Hamiltonians with sufficient accuracy to produce reliable quantum chemical insights. Finally, we extrapolate the requirements for future quantum hardware to enable practical and scalable quantum chemistry calculations using VQE algorithms. This work provides a roadmap for advancing quantum algorithms and hardware toward achieving quantum advantage in molecular modeling.}


\maketitle

\clearpage
\section{Introduction}\label{sec:intro}
Quantum computing holds the promise of solving problems that are intractable for classical computers, particularly in areas like cryptography, optimization, and material science.\cite{Feynman:1982} 
One of the most exciting prospects is its potential to revolutionize quantum chemistry\cite{Aspuru:2005,Cao:review:2019} and materials design by simulating complex quantum systems at a level of accuracy unattainable by classical methods.\cite{ibm:whitepaper:2024} 
By leveraging the principles of quantum mechanics, quantum computers can naturally represent the wavefunctions of molecular systems, offering a direct path to precise solutions for ground and excited-state energies, reaction mechanisms, and dynamic processes. This could lead to breakthroughs in drug discovery, sustainable energy solutions, and the development of advanced materials.

Within the Noisy Intermediate-Scale Quantum (NISQ) era, Variational Quantum Eigensolvers (VQEs)\cite{Peruzzo:2014} have emerged as one of the most promising algorithms for quantum chemistry applications.\cite{vqe_review:2022,Fedorov:vqe_review:2022} These hybrid quantum-classical algorithms are specifically designed to minimize the limitations of current quantum hardware, such as decoherence and gate errors, by delegating some computational tasks to classical processors.\cite{Preskill2018quantumcomputingi} 
VQEs utilize parameterized quantum circuits to approximate molecular ground states, with the optimization process guided by classical algorithms. 
Their adaptability and relative resilience to noise make VQEs particularly well-suited for investigating strongly correlated systems and capturing quantum effects critical to chemical properties.

However, employing VQE algorithms on real quantum hardware to solve intricate molecular Hamiltonians presents significant challenges. 
The limitations of current devices, including restricted qubit counts, high error rates, and limited coherence times, hinder the depth and accuracy of quantum circuits. 
Additionally, the need for repeated energy evaluations during the optimization process introduces computational overhead, exacerbated by measurement noise and sampling errors. 
These factors complicate the practical deployment of VQEs for large or strongly correlated systems, necessitating further algorithmic refinements and hardware advancements to fully unlock their potential.

This study investigates the capabilities and limitations of VQE algorithms implemented on current quantum hardware for electronic structure calculations of molecular systems. 
Our primary objective is to compute the ground-state energy, corresponding to the lowest eigenvalue of the time-independent Schr\"odinger equation, a fundamental quantity in quantum chemistry:  
\begin{equation}  
\hat{\mathcal{H}}\Psi = E\Psi  
\label{eq:schrodinger}  
\end{equation}  
To achieve this, we focus on one of the most promising VQE variants for molecular studies: the adaptive derivative-assembled pseudo-Trotter ansatz variational quantum eigensolver (ADAPT-VQE).\cite{Grimsley2019} 
We analyze multiple aspects of this methodology and propose various approximations, simplifications, and optimizations to enhance the performance of existing algorithms and their implementation.  
Finally, we identify the most computationally favorable scheme and use it to estimate the ground-state energy of the benzene molecule on a quantum computer. 
By doing so, we assess the current state of quantum hardware and determine how far it remains from enabling realistic and meaningful quantum chemistry calculations.  

The paper is organized as follows.
In Section~\ref{sec:results}, we introduce the overarching strategy employed to achieve our objectives. 
This is followed by a detailed discussion of key components necessary to enhance the performance of VQE-based approaches. 
Section~\ref{sec:hamiltonian} explores methods to simplify the electronic Hamiltonian, reducing computational complexity while maintaining accuracy. Section~\ref{sec:ansatz} focuses on optimizing the ansatz to balance expressivity and efficiency, and Section~\ref{sec:cobyla} addresses improvements in the numerical optimization of the parametric space. 
Section~\ref{sec:implementation} describes the design of an efficient implementation tailored for quantum computation.  
In Section~\ref{sec:quantum_experiment}, we integrate all the proposed optimizations and simplifications to compute the ground-state energy of benzene using an \texttt{IBM} quantum computer. 
Building on these results, Section~\ref{sec:coherence_requirements} evaluates the quantum hardware requirements necessary to obtain reliable results for our model system. 
Finally, the main findings of this work are resumed in the Conclusions section.

\section{Results and discussion}\label{sec:results}
To achieve the objectives of our study, we have designed a systematic procedure that optimizes and simplifies all aspects of the calculation, encompassing both the physical system and the algorithm, prior to performing the quantum experiment.
Our strategy begins by establishing a controlled, optimal scenario to test the VQE methodology under simplified conditions. 
This approach allows us to evaluate its fundamental capabilities. The performance observed in this scenario serves as a baseline for identifying limitations and extrapolating the methodology's applicability to more complex, real-world situations.
The procedure comprises the following key steps:
\begin{itemize}
    \item Simplifying the description of the physical system,
    \item Optimizing the VQE quantum algorithm, and
    \item Enhancing the implementation on a quantum computer.
\end{itemize}
These steps are addressed through the optimization and simplification of the molecular Hamiltonian, the structure of the wavefunction ansatz, and the minimization of statistical noise. 
Additionally, we employ various strategies to optimize the implementation of the generated quantum circuit on specific hardware platforms.
This systematic approach is applied to the calculation of the ground state energy of benzene using an \texttt{IBM} quantum computer, with the following goals:
\begin{itemize}
    \item Assessing performance under simplified conditions, and
    \item Extrapolating results to more realistic and challenging scenarios.
\end{itemize}
To illustrate our analysis, we present the results obtained for the benzene molecule, which serves as a representative model. 
Comparable performance has been observed for other systems, as detailed in the Supporting Information.

\subsection{Hamiltonian modeling} \label{sec:hamiltonian}

The complexity of the molecular electronic Hamiltonian arises primarily from the fact that, in principle, each electron interacts with all others. The second term in the rhs of equation~\ref{eq:h_total}, a two-body operator accounting for electron-electron interactions, is especially challenging and forms the foundation of an entire field dedicated to the development of approximate electronic structure methods,\cite{Blum_2024}
\begin{equation}
    \hat{\mathcal{H}} = \sum_{pq}{h_{pq} \hat a_p^\dagger \hat a_q} + \frac{1}{2} \sum_{pqrs}{g_{pqrs} \hat a_p^\dagger a_q^\dagger \hat a_r \hat a_s} \label{eq:h_total}
\end{equation}
where $h_{pq}$ and $g_{pqrs}$ are the one- and two-electron integrals, respectively, $\hat a^\dagger$ and $\hat a$ are the creation and annihilation operators, and $p$, $q$, $r$, $s$ are orbital indices.

\subsubsection{Active orbital space}
The number of terms in the Hamiltonian significantly influences the computational cost of VQE algorithms, as each term must be evaluated individually.
To simplify the complexity of the molecular electronic Hamiltonian, we employ the active space approximation. 
In this approach, the Hamiltonian is expressed by operators acting only on a selected subset of molecular orbitals, referred to as the active space.
This space is chosen to include orbitals that account for most of the ground-state correlation effects, typically those near the Fermi level. 
The remaining orbitals are divided into (i) core orbitals, i.e., fully occupied orbitals at lower energy, which are treated through an effective potential ($V_\text{eff}$), and higher-energy unoccupied orbitals, whose contributions are neglected.
By integrating out the contributions of the core electrons, the molecular Hamiltonian is reduced to an effective form:
\begin{equation} 
   \hat{\mathcal{H}}_\text{eff} = \sum_{pq}^\text{act}{\tilde h_{pq} \hat a_p^\dagger \hat a_q} + \frac{1}{2} \sum_{pqrs}^\text{act}{g_{pqrs} \hat a_p^\dagger \hat a_q^\dagger \hat a_r \hat a_s} + \hat{V}_\text{eff}, 
   \label{eq:h_eff}
\end{equation}
where the summations are restricted to the active space orbitals, $\tilde h_{pq}$ are the effective one-electron integrals and $\hat{V}_\text{eff}$ is the effective core potential operator arising from the frozen core orbitals (Supporting Information, Section~S1). 

In this study, we restrict the active space to four electrons in the two-fold $e_{1g}$ and $e_{2u}$ frontier $\pi$-orbitals, significantly simplifying the Hamiltonian.
Consequently, $\hat{V}_\text{eff}$ accounts for 38 inner electrons.
This reduced Hamiltonian captures the essential electronic interactions while dramatically lowering computational demands.

\subsubsection{Hamiltonian compression and orbital representation}
The size of the Hamiltonian is determined by the active space, the orbital basis, and the internal structure of the interactions between orbitals. 
A large Hamiltonian negatively impacts the precision of energy measurements due to the accumulated error in each individual term, which are measured separately. 
However, many of these terms may have negligible magnitudes. 
Removing such small terms is expected to have a minimal effect on the ground state energy while significantly reducing both the computational cost of the measurement and the associated error. 
This process is referred to as Hamiltonian compression.

Using the four-electron, four-orbital active space approach for the benzene molecule, the electronic Hamiltonian in equation~\ref{eq:h_eff} comprises 1057 terms. 
To alleviate the computational cost, we reduce the number of interacting elements to be evaluated by discarding one- and two-electron terms with negligible contributions, determined by setting an energy threshold. 
Retaining only terms with an absolute value greater than $10^{-2}$~Ha leads to a 78\% reduction in the number of Hamiltonian terms while maintaining the ground-state energy within a precision of approximately $10^{-3}$~Ha compared to the full calculation (Table~\ref{tbl:ham_compress}).\cite{integral_thresh}

Moreover, it has been demonstrated that using natural orbitals (NOs) derived from the unrestricted Hartree-Fock (UHF) density matrix, instead of canonical Hartree-Fock (HF) orbitals, can systematically enhance the convergence and overall performance of VQE algorithms for calculating molecular ground-state energies.\cite{Vaquero:VQEs:2024} 
This improvement is particularly notable in systems with strong electron correlation (entanglement). 
When expressed in the NO basis, the reduction in the Hamiltonian's dimensionality remains identical to that in the canonical HF basis (Table~\ref{tbl:ham_compress}). 
However, the resulting energy is consistently closer to the exact value, highlighting the advantages of employing NOs for VQE calculations.

\begin{table}[h]
   \centering
   \caption{Ground state exact (FCI) energy (in Ha) and total number of one- and two-electron terms in the molecular Hamiltonian (equation~\ref{eq:h_eff}) in the full and compressed orbital spaces.}
   \begin{tabular}{lccc}
   \hline
   \hline
    & full  & compressed MOs & compressed NOs  \\
    \hline
    Energy  & -227.94440224 & -227.94181062 & -227.94308835 \\
    \# terms   & 1057  & 233 & 233      \\
    \hline
    \hline
  \end{tabular}
  \label{tbl:ham_compress}
\end{table}

A detailed analysis of the number of Hamiltonian terms that must be evaluated for different active spaces, using both canonical and natural orbital representations, is provided in Section~S2 of the Supporting Information.  
These results demonstrate that the computational savings achieved through Hamiltonian compression become increasingly significant for larger active spaces.  
Furthermore, the superior performance of natural orbitals over canonical orbitals is consistently observed across various orbital spaces and compression thresholds.

\subsection{Ansatz optimization} \label{sec:ansatz}


The structure of the parameterized trial state is a crucial component of the VQE algorithm. 
It must represent a flexible unitary transformation that cannot be efficiently expressed on a classical computer while also addressing the optimization challenges associated with highly parameterized spaces.
Hardware-efficient ans\"atze, which are agnostic to the physical problem, have been argued to be unsuitable for accurately characterizing chemical systems.\cite{vqe_review:2022} 
Conversely, the constraints of current noisy quantum computers necessitate the use of compact ans\"atze to limit circuit depth.
Within the chemically inspired (or physically motivated) family of approaches, the ADAPT-VQE algorithm has demonstrated superior performance than other VQE methods. 
For instance, compared to the prototypical unitary coupled cluster (UCC) approach,\cite{Peruzzo:2014} ADAPT-VQE produces significantly more compact states,\cite{Grimsley2019} resulting in shallower quantum circuits.
ADAPT-VQE constructs the ansatz iteratively by adding the operator with the largest energy gradient at each step, 
\begin{equation}
    | \Psi \rangle  = \prod_k e^{ \theta_k \hat A_k } | \Psi_0 \rangle
    \label{eq:adapt_vqe}
\end{equation}
where $|\Psi_0\rangle$ denotes the initial state (typically Hartree-Fock), $\hat{A}_k$ is the Ferminonic anti-Hermitan operator added at the $k$-th iteration and $\theta_k$ the associated amplitude (optimized classically).
Furthermore, its iterative ansatz growth incorporates a parameter ``recycling" strategy, where optimal parameters from the previous step are reused. 
This approach accelerates convergence, mitigates the risk of local minima, and has proven particularly robust against barren plateaus.\cite{Grimsley:adapt:2023}

Specifically, applying ADAPT-VQE to calculate the ground-state energy of benzene (as detailed above) with a pool of single and double excitation operators requires a total of 6 iterations (operators) to achieve convergence with an error under chemical accuracy. 
This represents a 90\% reduction in the parameter space compared to UCCSD, which involves 64 operators.

\subsubsection{Symmetry constrains in the operator pool}
This reduction in the operator space is accompanied by an increase in the total shot count, which can significantly exceed that of UCC. 
This is primarily due to the fact that, at each ADAPT iteration, the number of energy gradients that need to be evaluated equals the total number of operators in the pool, a quantity that scales rapidly with system size. 
Therefore, in ADAPT-VQE, it is crucial to employ optimized operator pools that are as compact as possible while maintaining the necessary expressivity to accurately capture the features of the target state.
To reduce the size of the operator pool without compromising accuracy, one can leverage the symmetry properties of the target ground state, specifically its spin multiplicity and spatial symmetry. 
In our model case, the ground-state singlet of benzene belongs to the totally symmetric irreducible representation of the $D_{6h}$ point group (A$_{1g}$). 
Consequently, only excitation operators that preserve this symmetry, matching the symmetry of the initial state $|\Psi_0\rangle$, need to be included in the pool. 
Applying this symmetry restriction results in a 57\% reduction in the size of the single and double (occupied-to-virtual) excitation operator pool, i.e., from 14 to 6 operators. 

\subsubsection{Fermionic vs. Qubit operator pools}
Tang et al. have highlighted that state-preparation circuits generated by ADAPT-VQE remain too deep for practical use on current quantum devices.\cite{Tang2021} 
To address this, they introduced Qubit-ADAPT-VQE (or simply Qubit-ADAPT), a hardware-efficient variant of ADAPT-VQE. 
In Qubit-ADAPT, the ansatz is constructed directly using qubit-based operators rather than fermionic ones, with the qubit operator pool encompassing those present in the fermionic pool.
This guarantees that Qubit-ADAPT has potential to achieve the same level of accuracy as its fermionic counterpart.
The key advantage of Qubit-ADAPT lies in the significantly shallower circuits it produces, making the method far better suited for NISQ devices.

\begin{figure}[H]
\includegraphics[width=6.5cm]{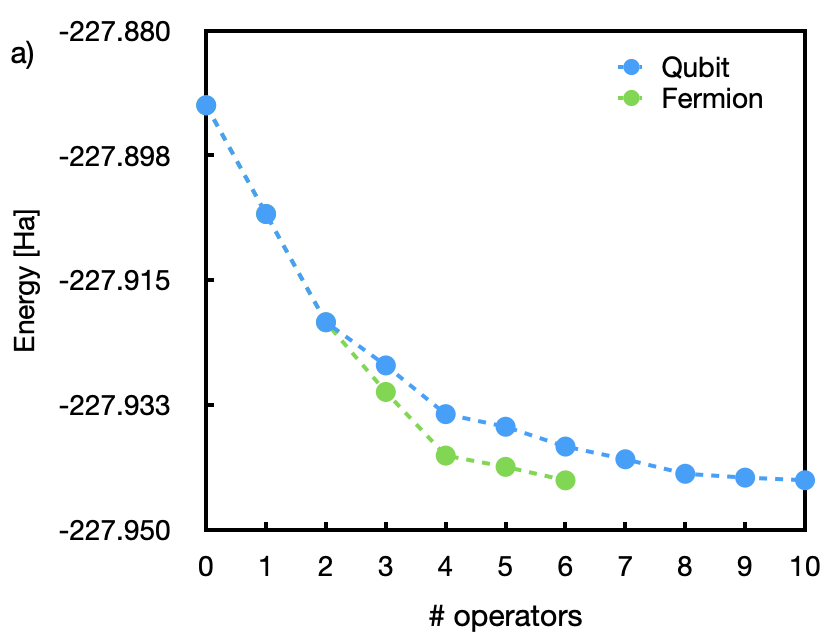}
\includegraphics[width=6cm]{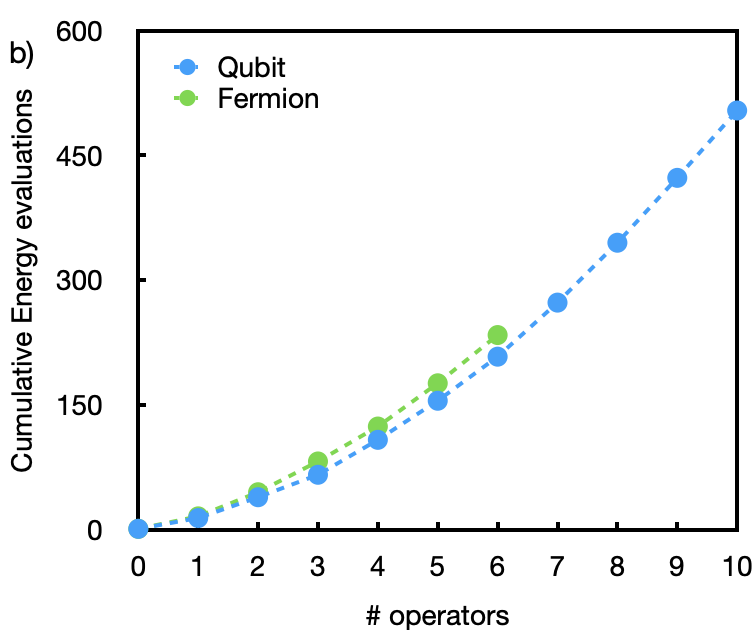}
\includegraphics[width=6.0cm]{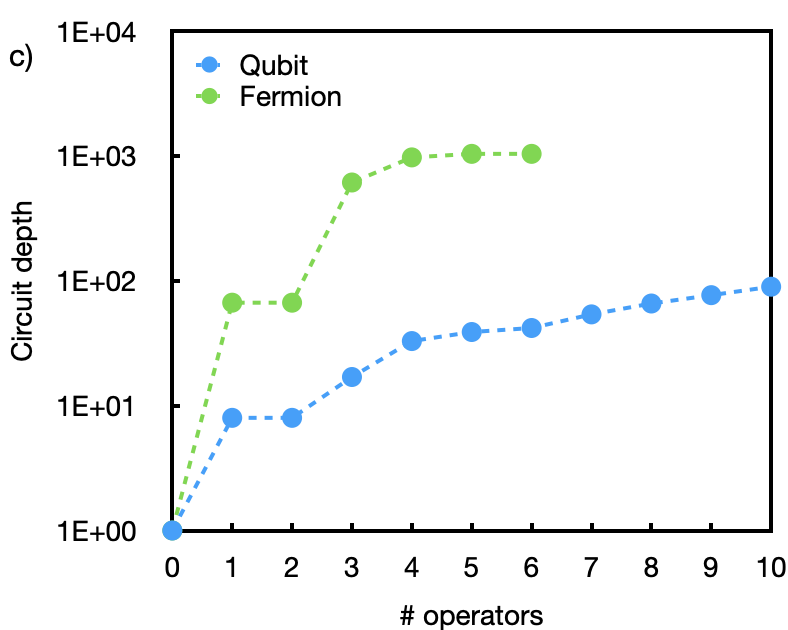}
  \caption{(a) Ground state energy, (b) number of energy evaluations, and (c) energy and gradient circuit depths as a function of the number of iterations for the simulation of the benzene molecule with fermion (green) and qubit (blue) ADAPT-VQE algorithm.}
  \label{fig:vqe_comparison}
\end{figure}

Figure~\ref{fig:vqe_comparison} presents a performance comparison between ADAPT-VQE using fermionic and qubit operators. 
Fermionic operators require a smaller number of iterations to achieve convergence (Figure~\ref{fig:vqe_comparison}a), resulting in fewer energy evaluations (Figure~\ref{fig:vqe_comparison}b).
However, each added fermionic operator significantly increases the size of the ansatz compared to the qubit pool. 
Consequently, the circuits generated to evaluate the ground state energy and the energy gradients by Qubit-ADAPT at each iteration are substantially shallower than those constructed with fermionic operators (Figure~\ref{fig:vqe_comparison}c). 
Notably, the maximum circuit depth required for the energy evaluation (at the final iteration before convergence) is reduced by an order of magnitude in Qubit-ADAPT.
Interestingly, we observe that in the fermionic pool the cost of the energy evaluation is much larger than the cost of the gradient evaluation, while in the case of the qubit pool the circuit depths for the energy and gradient are of similar magnitude. 

These results, however, do not fully reflect the impact of the larger size of the qubit operator pool compared to the fermionic pool. 
While the cost of gradient evaluations is generally lower than that of energy evaluations, the significantly larger number of operators in the qubit pool necessitates evaluating more gradients at each ADAPT iteration. 
Consequently, the total number of CNOT gates required across all circuits throughout the full fermionic or qubit simulations is nearly identical (Figure~S6). 
Nonetheless, this number remains substantially smaller than the total count of two-qubit gates employed solely for energy evaluations. 
As a result, the combined computational cost for Qubit-ADAPT remains lower than that for fermionic ADAPT.
Moreover, the consistently shallower circuit depths generated in Qubit-ADAPT provide a crucial advantage, enhancing its resilience to quantum noise and making it better suited for implementation on NISQ hardware.
Therefore, Qubit-ADAPT will be employed through the rest of the study.

\subsection{Minimizing statistical noise} \label{sec:cobyla}

The hybrid nature of VQE methods relies on optimizing the parameter space through a classical optimizer, making this component essential for the algorithm's performance.
In classical optimization tasks, gradient-based optimizers typically perform well when the potential energy landscape is smooth and contains few local minima. \cite{claudino_benchmarking_2020}
However, in the context of VQE, energy evaluations (especially on current NISQ devices) are subject to inherent noise and have a standard error associated with it. Consequently, optimizers that do not rely on gradients often yield better results. \cite{singh_benchmarking_2023} 
In this sense, the COBYLA (Constrained Optimization BY Linear Approximations)\cite{Powell:cobyla:1994} algorithm has become a preferred choice for VQE.\cite{McClean_2016} 
COBYLA iteratively approximates the objective function within a trust region using linear models, optimizing the parameters without requiring explicit derivative information. 
Its simplicity and ability to handle noisy function evaluations make it particularly suited to VQE applications.
Despite these advantages, COBYLA can struggle with VQE simulations where energy sampling is performed with a low number of shots. Under such conditions, it may converge to undesired local minima and require an excessive number of iterations to attempt escaping them, often without success.
To address the limitations of COBYLA in the context of Qubit-ADAPT, we propose two tailored modifications: (i) parameter pre-scan and (ii) dynamic tolerance criterion.

In (Qubit-)ADAPT-VQE, parameter optimization occurs iteratively, using the values optimized in the previous iteration as initial guesses for subsequent updates. 
This iterative recycling of parameters is a critical factor in the algorithm’s success.\cite{Grimsley:adapt:2023} 
However, a common challenge arises when optimizing the parameter associated with the most recently added operator, which typically starts with a value of zero. This initialization can hinder efficient convergence.
In the specific case of Qubit-ADAPT, each exponential operator consists of two components: a Pauli string and a coefficient. 
The Pauli string is constructed as a direct product of normalized Pauli operators. 
As a result, the ansatz exhibits a periodicity of $2\pi$ with respect to variations in the coefficient (Figure~S8).
Notice that in fermion ADAPT-VQE, the mapping to the qubit basis followed by Trotterization introduces a prefactor coefficient that alters the periodicity of each operator.
Leveraging the periodic structure in Qubit-ADAPT, we initialize the parameter associated with the most recently added operator in the ansatz by performing a scan over the range $-\pi$ to $\pi$, while keeping all other parameters fixed at their previously optimized values.
The parameter value that minimizes the cost function in this range is then used as the initial guess for the COBYLA optimization. 

On another note, the tolerance parameter in COBYLA plays a critical role in defining the convergence criteria and directly influences both the number of optimization steps and the overall computational cost. 
In fault-tolerant devices, such as classical computers, the precision of energy evaluations far exceeds the specified tolerance criteria, making its impact on the optimization negligible. 
However, on current NISQ quantum devices, the standard error associated with energy measurements can surpass the predefined COBYLA tolerance value. 
This mismatch often leads to an unnecessary increase in optimization steps, thereby inflating the computational cost.
To address this issue, we introduce a second modification to COBYLA: a dynamic adjustment of the convergence tolerance parameter based on the standard error of energy measurements. 
Specifically, following the pre-optimization procedure described earlier, if the initially defined tolerance is found to be smaller than the standard error associated with the initial guess, the tolerance is updated to match this error estimate. 
This approach ensures that the optimization process remains efficient while accounting for the inherent noise in NISQ devices.

\begin{figure}[H]
  \includegraphics[width=6.4cm]{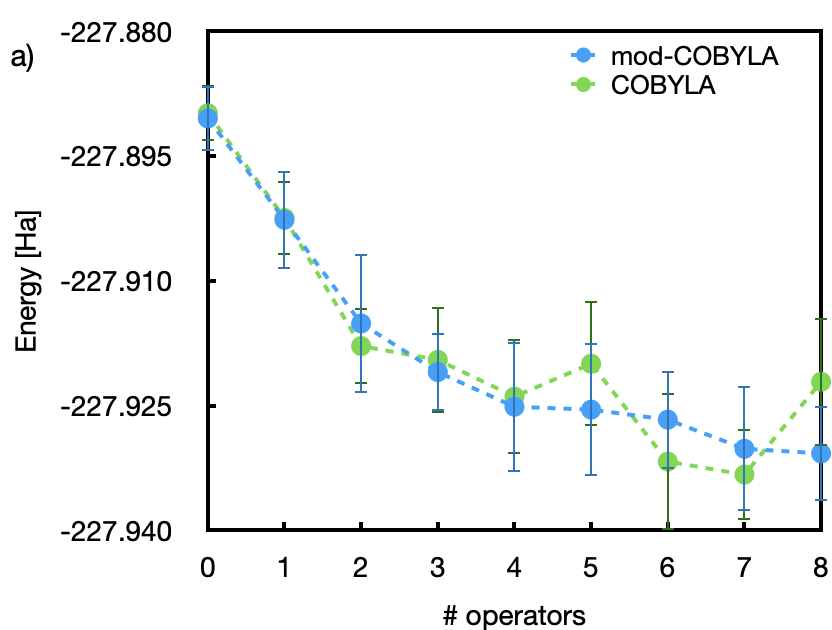}
  \includegraphics[width=6cm]{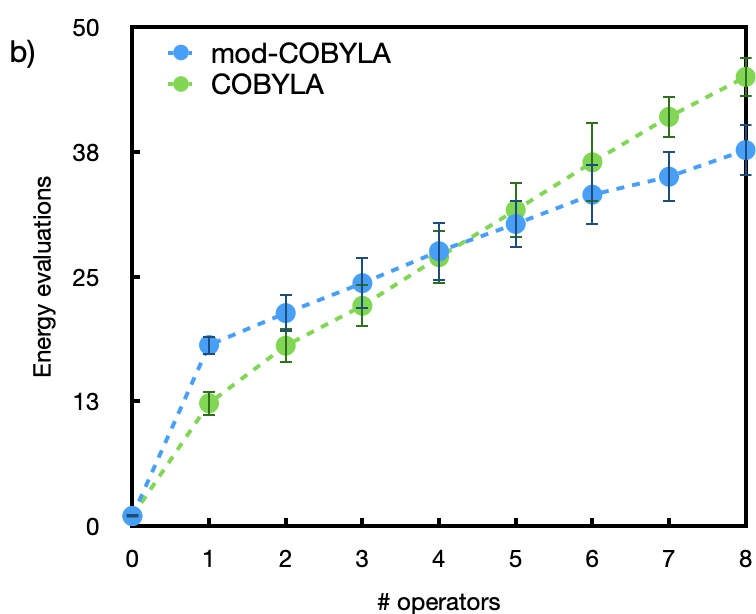}
  \caption{(a) Ground state energy and (b) energy evaluations of benzene molecule along the qubit-ADAPT iterations using COBYLA (green) and Mod-COBYLA (blue). Each point corresponds to the average of 30 simulations using 500 shots. Default tolerance in standard COBYLA is $10^{-3}$~Ha.}
  \label{fig:opt_500}
\end{figure}

Figure~\ref{fig:opt_500} presents the results of a Qubit-ADAPT simulation for the benzene molecule. This simulation was conducted using the \texttt{Qiskit} library to generate quantum circuits and the \texttt{Aer simulator} for energy evaluations, employing 500 shots per measurement. 
The plots compare the performance of the standard COBYLA optimizer, as implemented in \texttt{scipy} with its default tolerance of $10^{-3}$~Ha, against our modified COBYLA approach, which we refer to Mod-COBYLA. 
The modifications include the parameter pre-scan strategy and a dynamic tolerance adjustment based on the standard error of energy measurements.

Comparing the performance of standard COBYLA with Mod-COBYLA, we observe a similar rate of energy convergence. 
However, Mod-COBYLA demonstrates better scaling with the number of energy evaluations required during the simulation. 
This improvement highlights how dynamically adjusting the tolerance value prevents the optimizer from performing unnecessary additional evaluations. 
The benefits of this approach become more pronounced when the number of shots is small, and the standard error of the energy measurement is relatively large (Figures~S9 and~S10).
Similar results are observed in the ground-state energy evaluations of other molecular systems. 
In fact, the advantages of Mod-COBYLA when combined with Qubit-ADAPT become even more pronounced when characterizing strongly correlated systems (see Section~\ref{sec:cobyla_H4} in the Supporting Information).

\subsection{Efficient implementation in a quantum hardware} \label{sec:implementation}

\subsubsection{Divide and conquer strategy} 
Next, we intend to implement the streamlined and optimized approaches outlined above to compute the ground state energy of benzene on a quantum device.
Existent \texttt{IBM} quantum computers impose limits on the job size that can be run, including restrictions on circuit depth, the number of observables to be measured, and the level of error mitigation applied during calculations.
On average, at the time of writing, this limit ($L$) is defined by the product of the number of two-qubit gates and the number of observables being $L\approx36000$ according to our tests (S6).
Since quantum circuits cannot be arbitrarily divided, we address this constraint by splitting the Hamiltonian into smaller subsets of observables, allowing each subset to be evaluated in separate jobs. 
Each energy evaluation therefore requires executing $n_j$ separate jobs:
\begin{equation}
n_j = \ceil*{\frac{d~n_{H}}{L}} 
\end{equation}
where $\ceil{~}$ denotes the ceiling function, $d$ is the total depth of the transpiled circuit, $n_H$ is the number of terms in the Hamiltonian, and $L=33000$ is chosen as a conservative threshold to ensure that each job can be executed within the hardware limits.
In the specific case of benzene calculations using a compressed Hamiltonian ($n_H = 233$ terms) and the Qubit-ADAPT protocol, we found that the entire Hamiltonian could be evaluated within a single job ($n_j = 1$ for ans\"atze containing up to 3 operators), or split into two jobs ($n_j = 2$ for ans\"atze with 4 to 7 operators).
This approach ensures that even the deepest circuits with error mitigation can be executed efficiently.

\subsubsection{Circuit optimization}
\begin{figure}[H]
  \includegraphics[width=13cm]{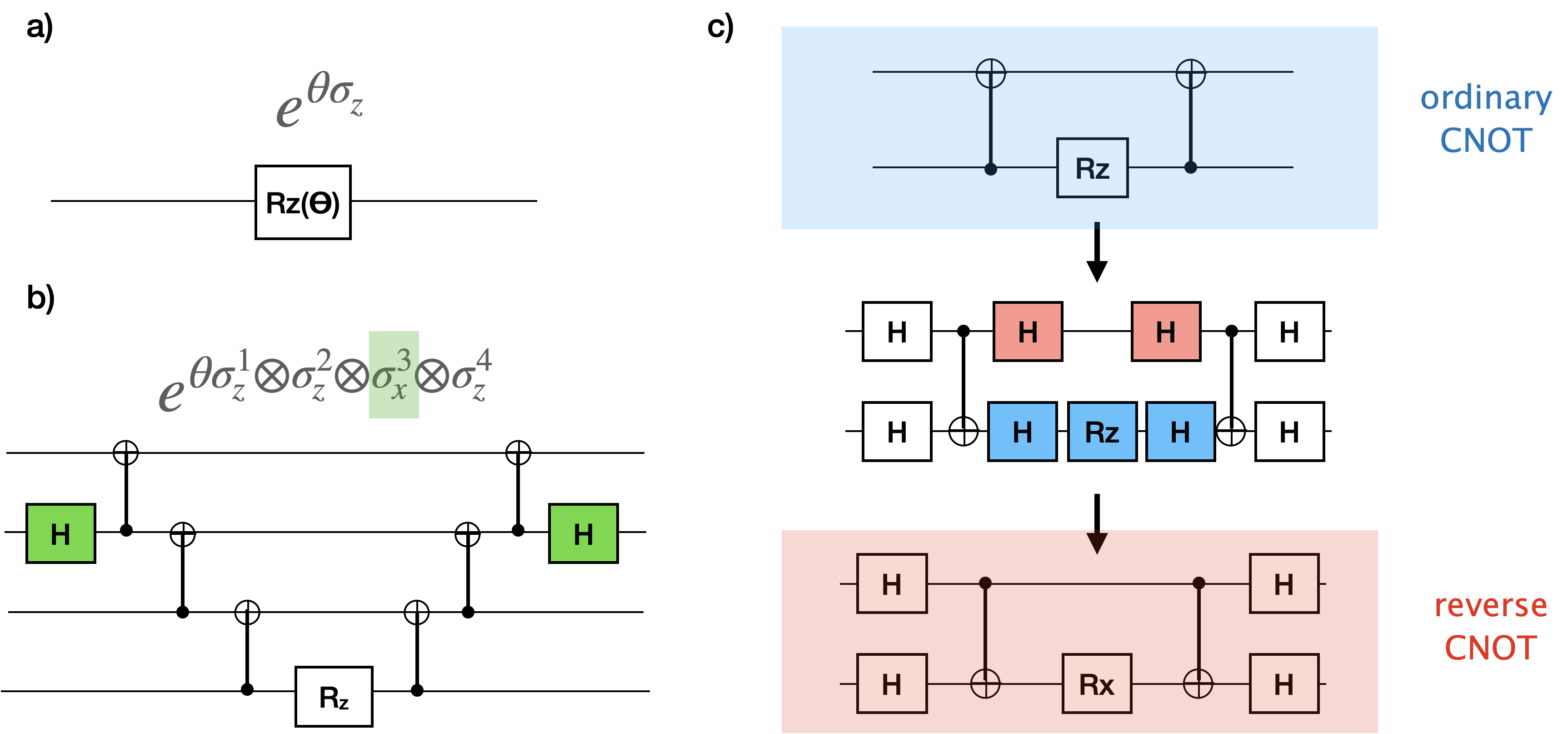}
  \caption{Circuit corresponding to the staircase algorithm for (a) a single exponential operator with one Pauli operator $\sigma_z$, and (b) a Pauli string ($\sigma_z^1\otimes\sigma_z^2\otimes\sigma_x^3\otimes\sigma_z^4$), with superindices referring to the qubit index. 
  (c) Two different orientations of the CNOT gates in the staircase algorithm: standard and reverse.}
  \label{fig:staircase}
\end{figure}

Reducing the depth of quantum circuits is crucial to obtaining accurate results, especially given the noise inherent to current quantum hardware. 
In this work, the ansatz is implemented using the staircase algorithm,\cite{mansky_decomposition_2023} a structured decomposition method that systematically constructs the exponential operator, as illustrated in Figure~\ref{fig:staircase}a,b.
This algorithm consists of a cascade of CNOT gates surrounding a Z-rotation gate that encodes the scalar parameter $\theta$ of the operator. To account for the various $\sigma_i$ operators in the Pauli string, additional single-qubit gates are applied to each qubit.

The CNOT gates in the staircase algorithm can be implemented in two possible orientations: standard and reverse.\cite{mansky_decomposition_2023} 
The key difference between these orientations lies in the Hadamard gates surrounding the CNOTs. 
Due to the structure of these circuits, these Hadamard gates can often be canceled with adjacent gates, resulting in an alternative, shallower equivalent circuit. 
The choice between standard and reverse orientations is therefore dictated by the specific Pauli string of the exponential operator being implemented, with one orientation typically offering a more compact circuit depth than the other.

\begin{figure}[H]
  \includegraphics[width=6.4cm]{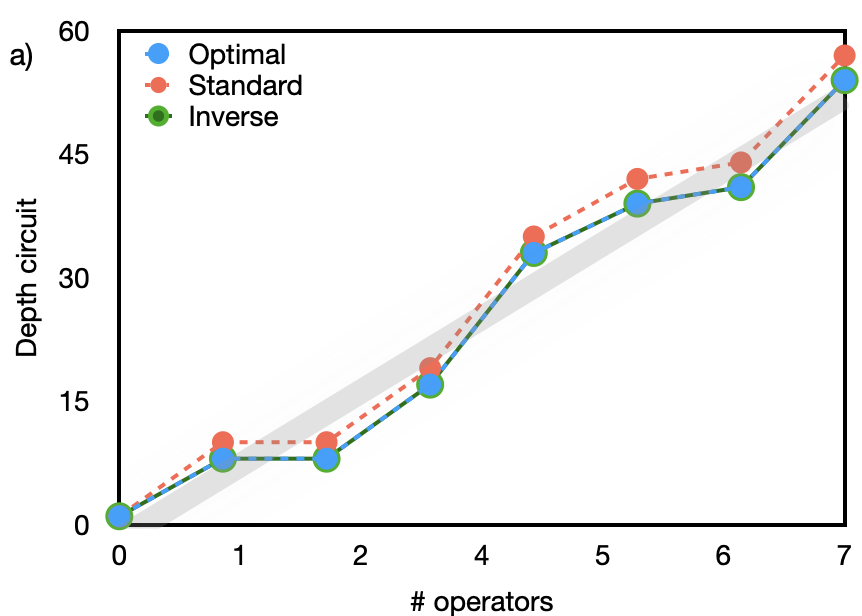}
  \includegraphics[width=6.4cm]{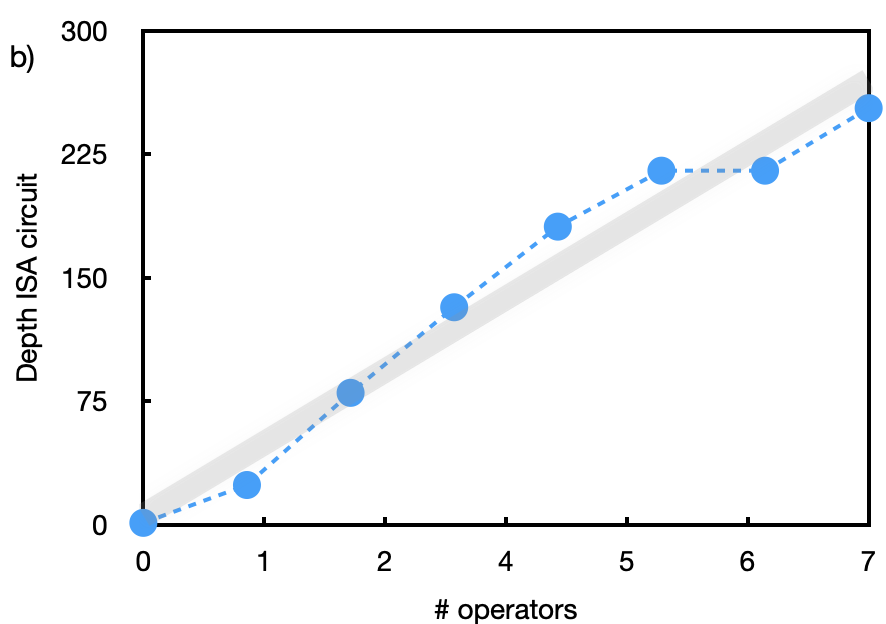}
  \caption{Circuit depth obtained with standard and optimized staircase algorithms (a), and ISA depth (b) as a function of the number of qubit operators.}
  \label{fig:qc_counts}
\end{figure}

To optimize the quantum circuits generated by Qubit-ADAPT, and more generally, any VQE implementation based on the staircase entangling pattern, we have developed an algorithm that determines the optimal orientation of each CNOT gate to minimize the overall circuit depth.  
Our strategy exploits the symmetry of the staircase algorithm to cancel adjacent gates, shortening the overall circuit.
Figure~\ref{fig:qc_counts}a compares the circuit depths obtained using the standard and optimized implementations, showing that the latter systematically produces shallower circuits. 
In the present case, where the total number of CNOT gates is relatively small, the optimized implementation yields circuit depths comparable to those obtained with a simple reverse orientation of the staircase pattern (Table~S1).  
However, for deeper circuits, particularly those generated by fermionic ADAPT-VQE in this work (Table~S2), our optimization strategy outperforms both the standard and reverse orientation approaches, consistently leading to more compact circuits.

However, these circuits are not directly executable on the hardware, as the specific constraints of the target quantum processor, such as the native gate set, qubit connectivity, and error mitigation strategies, must be taken into account.
The process of translating the designed, hardware-agnostic circuit into a form compatible with the available machine instructions is known as transpilation. 
This step decomposes the original gates into the device’s native gates and inserts additional operations (e.g., SWAP gates) to account for limited qubit connectivity.
Here, we transpile the optimized Qubit-ADAPT circuits using the \texttt{Qiskit} library, selecting optimization level 3 to reduce circuit depth. 
We observe that the depth of the transpiled circuits, referred to as the Instruction Set Architecture (ISA) depth, is approximately five times greater than that of the original logical circuit. 
This five-fold overhead primarily originates from the way CNOT gates are implemented on the \texttt{IBM-Torino} quantum processor, where limited qubit connectivity requires additional SWAP gates and gate decompositions, ultimately increasing the total circuit depth.
Moreover, this depth increases nearly linearly with the number of ADAPT iterations (Figure~\ref{fig:qc_counts}b).

\subsubsection{Layout optimization}
Optimizing the qubit layout is essential for executing quantum circuits efficiently and maximizing measurement precision. 
One of the primary sources of error in quantum computations arises from qubit decoherence, which varies across different qubits on a quantum chip.  
Additionally, the structure of quantum circuits based on the staircase algorithm predominantly involves CNOT gates that connect neighboring qubits. 
Consequently, a layout with linear, consecutive qubits is particularly well-suited for these circuits.
To address this challenge, we have developed an automated procedure that systematically identifies and selects the optimal qubit layout simultaneously accounting for both qubit decoherence times and connectiviy.

To determine the optimal qubits on the hardware chip, we first retrieve information about the readout error and the decoherence times ($T_1$ and $T_2$) for each qubit. Using this data, we compute a quality function $f_q$ defined as:
\begin{equation}
    f_q= \sum_{i=1}^{N_q} \left\{ \frac{1 - \epsilon_2^i}{1 - \epsilon_2} + \frac{T_1^i}{T_1} + \frac{T_2^i}{T_2} \right\}
    \label{eq:qubit_quality_func}
\end{equation}
where $N_q$ is the number of qubits, $\epsilon_2^i$ corresponds to the two-qubit gate error and $\epsilon_2$, $T_1$ and $T_2$ indicate average values among all qubits. 
Using this quality function, we select the set of consecutive qubits that maximizes $f_q$, ensuring the best layout for the circuit. 

\begin{figure}[H]
  \includegraphics[width=6.6cm]{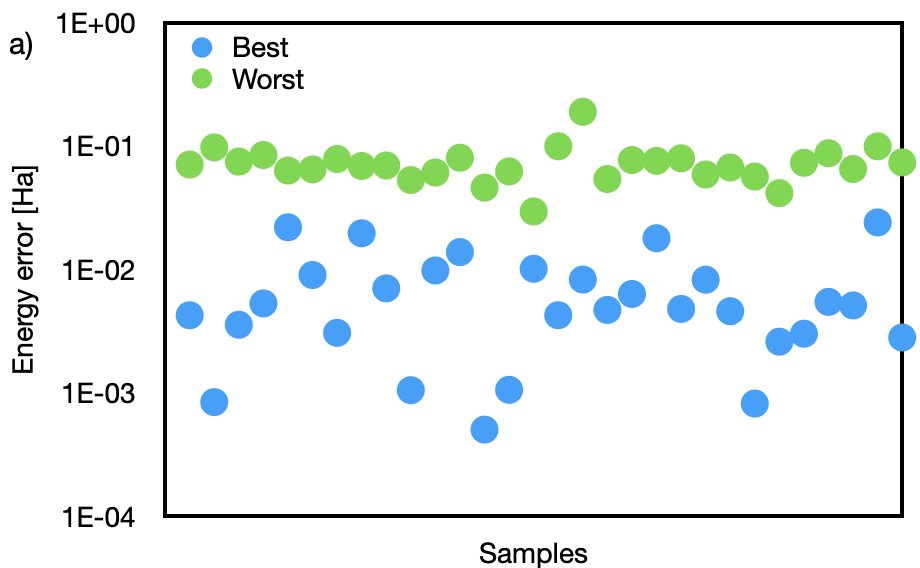}
  \includegraphics[width=6.4cm]{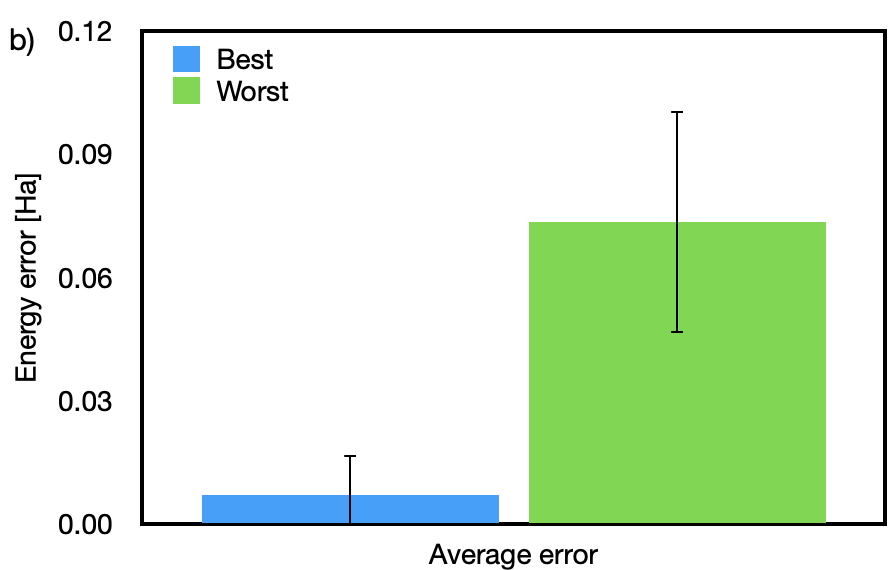}
  \caption{Energy errors (in Ha) of 30 individual experimental measures (a) and in average (b) in the evaluation of benzene ground state energy calculated using one exponential operator with the best and worst layout.
  Vertical bars in (b) indicate deviation error bars.
  The energy error is computed as the evaluated energy using \texttt{IBM-torino} with 9000 shots with respect to the exact energy computed in a classical computer.}
  \label{fig:layout}
\end{figure}

To test and validate this procedure, we evaluate the energy of the Qubit-ADAPT ansatz for the benzene molecule after the first iteration, that is, using a single exponential operator.  
This example was chosen because the corresponding ISA-circuit has a depth of 24, making it shallow enough to achieve high-precision energy estimates reliably. 
This allows us to systematically assess the impact of different qubit layouts on the accuracy and efficiency of the computation.
Figure~\ref{fig:layout} compares the performance of the best and worst qubit layouts, as ranked by our quality function $f_q$ (equation~\ref{eq:qubit_quality_func}).
The results, obtained from 30 independent energy evaluations for each layout, reveal a striking performance gap: energy errors are systematically and significantly larger when using the qubit layout with the lowest $f_q$ value.
On the other hand, the average error obtained with the optimal layout is approximately an order of magnitude smaller than that of the worst-performing layout.
These results emphasize the critical importance of qubit layout optimization for achieving accurate energy estimations in VQE simulations.

\subsection{Quantum hardware experiment} \label{sec:quantum_experiment}

To assess the performance of our Qubit-ADAPT approach on quantum hardware, we compute the ground state energy of benzene at each iteration, i.e., after the addition of each qubit operator.  
Given the high error rates in current quantum devices, the quantum machine is used exclusively for single-point energy evaluations, while all other computational steps, including gradient-based operator selection and amplitude optimization using Mod-COBYLA, are carried out via classical simulations.  
In other words, at each iteration, we construct the optimized ansatz from exact calculations and evaluate its energy on the quantum computer.  
These experiments were conducted on the \texttt{IBM-Torino} machine, 
incorporating error mitigation techniques available in \texttt{Qiskit}, specifically zero noise extrapolation (ZNE) and twirled readout error extinction (TREX).
To assess the reliability of the results, we performed 30 independent runs per iteration, each consisting of 9000 measurement shots.

\begin{figure}[H]
    \centering
    \includegraphics[width=6.5cm]{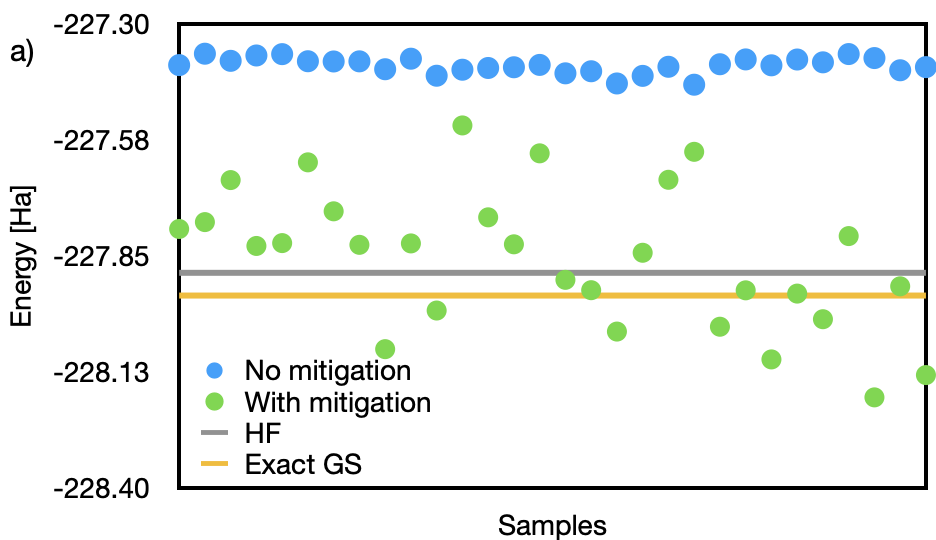}
    \includegraphics[width=5.7cm]{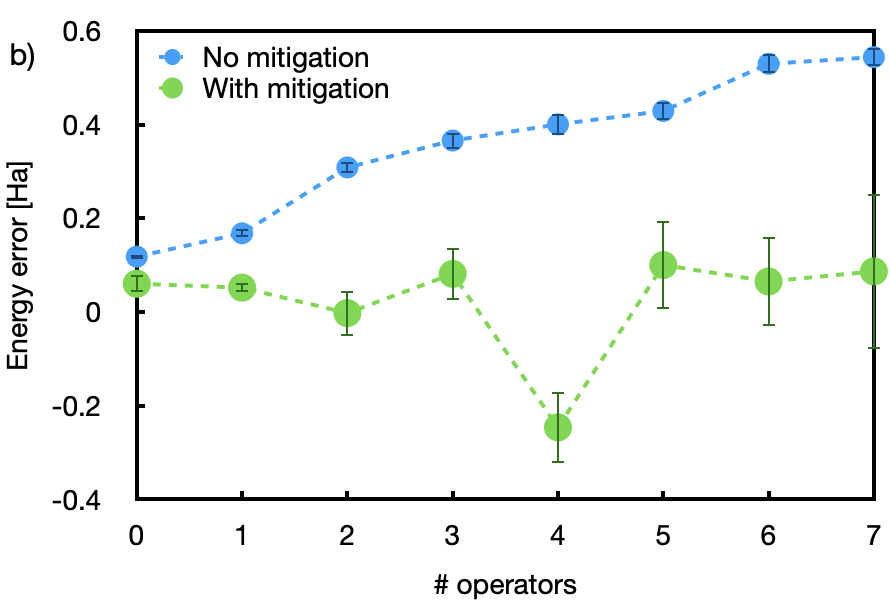}
    \caption{(a) Energy evaluations (in Ha) for the Qubit-ADAPT ansatz with 7 qubit operators.
    (b) Error with respect to analytical ansatz energy (in Ha) for the Qubit-ADAPT ans\"atze with 0-7 operators of benzene ground state with \texttt{IBM-Torino} (9000 shots) with (green) and without (blue) error mitigation. }
    \label{fig:stats_rh}
\end{figure}

Figure~\ref{fig:stats_rh}a presents the energy evaluations for the 7-operator Qubit-ADAPT ansatz.  
Despite the various simplifications and approximation strategies employed, resulting in relatively shallow circuits, the results are rather disappointing. 
The large errors stem from the extensive number of measurements required to evaluate the full Hamiltonian, a direct consequence of the complexity and size of electronic structure Hamiltonians.
As a result, most measurements exhibit significant deviations from the exact Qubit-ADAPT energy, often exceeding the total correlation energy range, i.e., the energy difference between the analytical HF and Qubit-ADAPT results.  
For completeness, the results for ans\"atze containing 0 to 6 qubit operators are shown in Figure~\ref{fig:samples_rh}.

The overall results and statistical analysis for ans\"atze containing 0–7 operators, with and without error mitigation, are summarized in Figure~\ref{fig:stats_rh}b.
Without error mitigation, energy estimates consistently lie above the exact Qubit-ADAPT solutions, with errors growing rapidly as the number of operators increases. 
Moreover, the standard deviation across independent samples remains relatively small, indicating stable but biased estimations.
In contrast, error mitigation using ZNE significantly reduces the average energy error but introduces a substantial increase in variance. 
This behavior, lower bias at the expense of higher variance, is a well-known feature of ZNE techniques.\cite{GiurgicaTiron2020}
Unfortunately, the increased variability in the mitigated results leads to outlier instances where individual energy estimates display very large errors, despite the application of error mitigation.
In all cases, both with and without mitigation, the standard deviation increases with the size of the ansatz, reflecting the growing sensitivity to noise as circuit depth increases. 
This trend is expected, as larger ans\"atze generate more complex quantum circuits that amplify the impact of hardware imperfections.
Overall, these results underscore that, given the magnitude of the observed errors in molecular energy evaluations, reliable VQE optimization on current quantum hardware remains unfeasible.

\subsection{Coherence time requirements} \label{sec:coherence_requirements}

Building on the limitations discussed in the previous section regarding the implementation of Qubit-ADAPT, we now turn to evaluating the algorithm’s error tolerance by investigating the impact of noise on its performance.
To this end, we conducted a series of simulations using a thermal relaxation noise model from the \texttt{Qiskit Aer} simulator. 
In these simulations, energy evaluations were performed at each Qubit-ADAPT iteration, with operator coefficients obtained from exact classical calculations.
To systematically assess the effects of decoherence, we analyzed how the computed energy varies as coherence times are progressively increased. 
Specifically, we scaled the relaxation and dephasing times as \(T_1 = \alpha T_1^{(0)}\) and \(T_2 = \alpha T_2^{(0)}\), where \(T_1^{(0)} = 100\) $\mu s$ and \(T_2^{(0)} = 150\) $\mu s$ represent typical values for current quantum hardware, and \(\alpha \geq 1\) is a scaling factor controlling the noise level.
For these simulations, we follow a procedure similar to that outlined in the \texttt{Qiskit} documentation.\cite{qiskit_noise_models}  
The execution times for specific quantum operations, including gate applications, qubit resets, and measurements, are summarized in Table~S3.
   
\begin{figure}[H]
\centering
  \includegraphics[width=6.35cm]{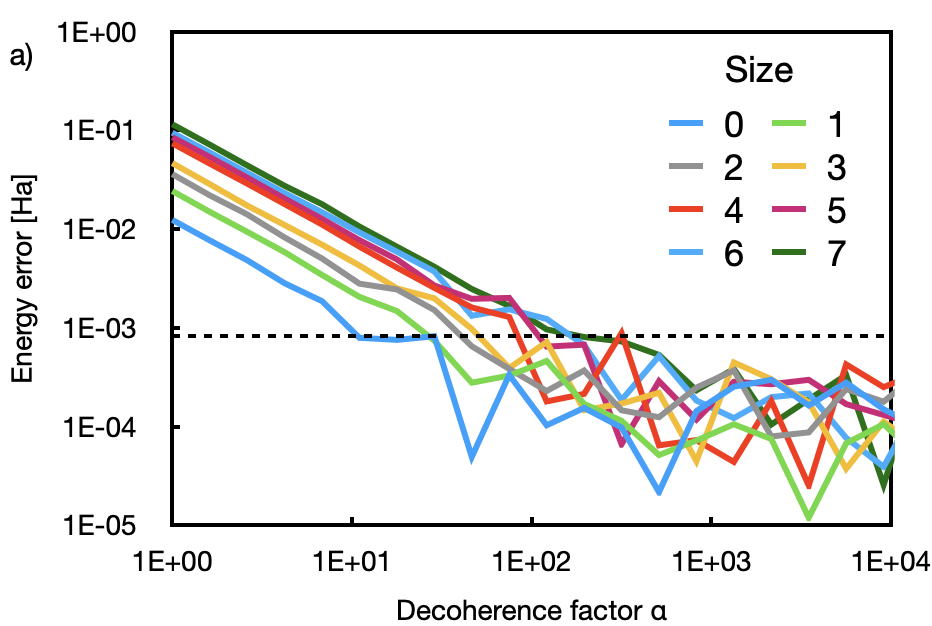}
  \includegraphics[width=6cm]{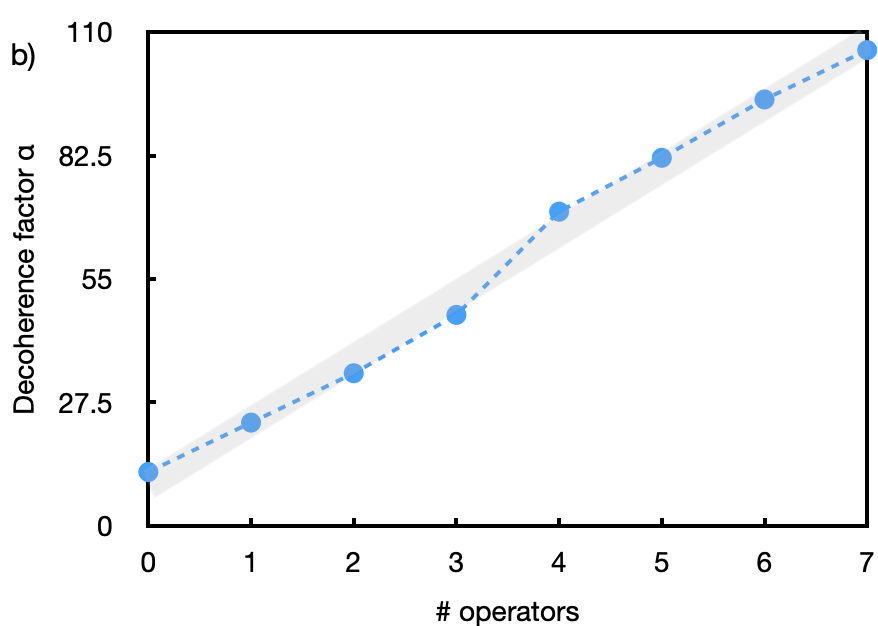}
  \caption{(a) Energy error (horitzontal line at 1 mHa) (in Ha) of Qubit-ADAPT ans\"atze with different number of excitation operators for the ground state of benzene as a function of the noise scaling factor ($\alpha$).
  Simulations carried out using the \texttt{Qiskit Aer} simulator.
  Each data point corresponds to the average of six computed replicates obtained with 9000 shots each. 
  (b) Decoherence factor necessary to achieve errors smaller than 1 mHa as a function of the number of qubit operators [extrapolated from (a)].
  }
  \label{fig:noise_model}
\end{figure}

Figure~\ref{fig:noise_model}a shows the average energy error as a function of the noise scaling factor ($\alpha$) for Qubit-ADAPT ans\"atze including between 0 and 7 excitation operators. The energy error exhibits an approximately exponential decay with increasing decoherence, which appears linear on the logarithmic scale. Below 1 mHa, the curves display irregular behavior due to sampling noise and finite-precision effects.

For the typical coherence times considered ($\alpha = 1$), noise effects are excessively large (too short $T_1$ and $T_2$ times), leading to computed energies that are significantly overestimated compared to the exact values.  
As the number of operators increases, the corresponding circuits become deeper, further amplifying these errors.
Consequently, ans\"atze with more variational parameters exhibit larger total energies, in contradiction with the variational principle, highlighting the detrimental impact of noise on Qubit-ADAPT performance.
This behavior is consistent with the energy values obtained from quantum hardware experiments and enables us to extrapolate the required coherence times necessary to achieve reasonable energy precision for Qubit-ADAPT evaluations on real hardware. 
Reducing noise, that is, increasing coherence times to perform quantum operations, gradually mitigates energy overestimations, recovering the expected energy ordering between ans\"atze (lower energies for ans\"atze with more operators) and bringing results closer to exact values.
Our analysis suggests that achieving energy errors below 1 mHa relative to  analytical values would require decoherence times growing almost linearly with the number of qubit operators (Figure~\ref{fig:noise_model}b). 
Specifically, obtaining accurate Qubit-ADAPT energies for ans\"antze with up to seven operators (iterations) would necessitate coherence times approximately two orders of magnitude longer than those currently available ($\alpha\approx100$).

However, these predictions only refer to the energy evaluation of Qubit-ADAPT ans\"atze and do not account for the full optimization process, which involves gradient evaluations and parameter optimization.
To assess the impact of noise on the complete VQE optimization procedure, we performed multiple Qubit-ADAPT simulations using the \texttt{Qiskit Aer} simulator with 1000 shots per measurement. 
A noise model with varying decoherence times was employed to systematically investigate its effect.
Notably, in these simulations, the convergence criteria were intentionally removed, allowing the algorithm to run indefinitely and providing a clearer picture of how noise influences the optimization trajectory.
For each noise configuration, we conducted 30 independent replicates and computed the average energy to capture statistical variations and trends.

\begin{figure}[H]
  \includegraphics[width=6.3cm]{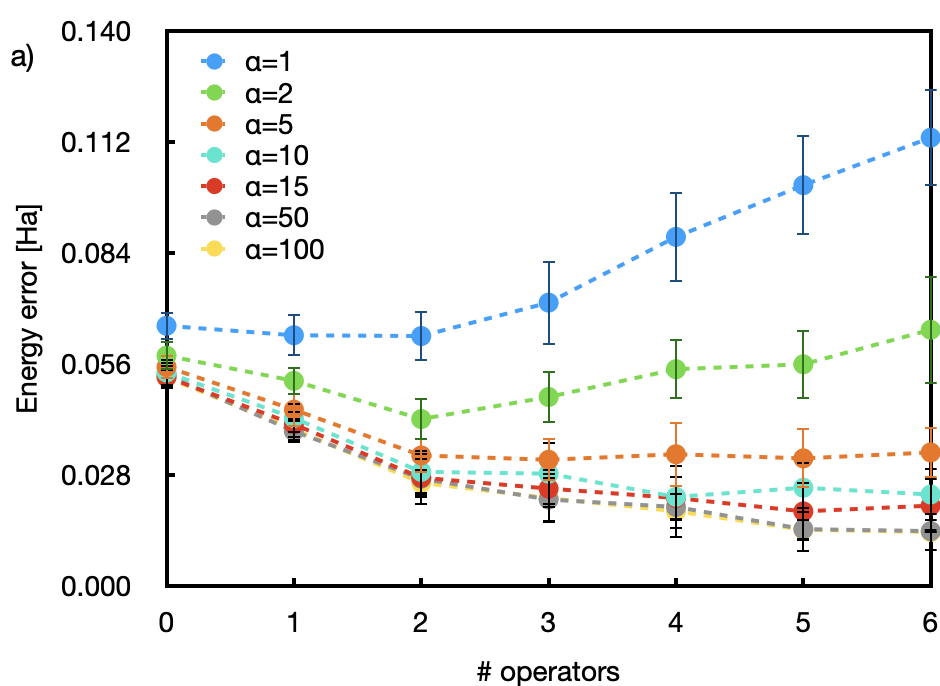}
  \includegraphics[width=6.4cm]{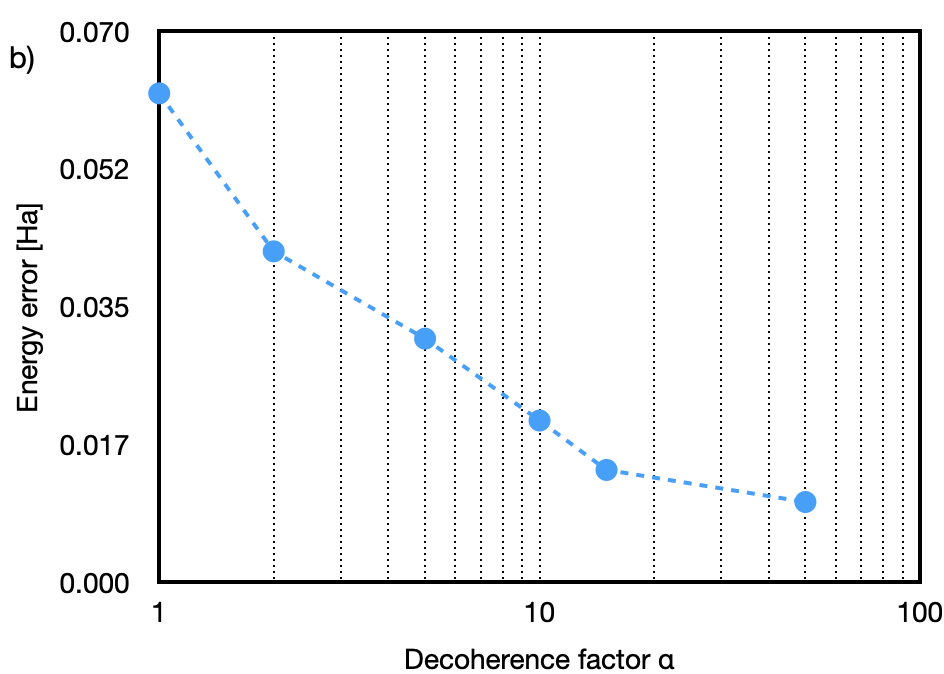}
  \caption{(a) Energy errors along ADAPT iterations (a) and as a function of the coherence parameter $\alpha$ (b) of full Qubit-ADAPT simulations with thermal relaxation noise for the ground state of benzene. 
  Values in (b) correspond to the lowest errors in panel (a).
  All errors given with respect to the analytical Qubit-ADAPT energy with eight operators.}
  \label{fig:noise_vqe}
\end{figure}

The results of the full Qubit-ADAPT simulations are presented in Figure~\ref{fig:noise_vqe}.  
For short coherence times ($\alpha < 5$), the simulations fail to variationally converge toward the analytical solution.  
As the number of iterations increases, cumulative noise errors compound due to the growing circuit depth, eventually reaching a point where they become too large to overcome.  
Consequently, instead of progressively lowering the energy as expected, the optimization breaks down, leading to increasing energy errors (Figure~\ref{fig:noise_vqe}a).
Increasing coherence times helps stabilize energy errors; however, achieving energy accuracies on the order of 1~mHa in noisy simulations requires coherence times to be scaled by a factor of approximately 50–100 (Figure~\ref{fig:noise_vqe}b).

\section{Conclusions}\label{sec:conclusion}
In this work, we have explored the application of VQEs to determine the ground state of molecular electronic structure Hamiltonians on current quantum hardware, using the benzene molecule as a prototypical medium-sized system. We addressed the key limitations of NISQ devices by implementing a comprehensive strategy involving Hamiltonian simplification, ansatz design, classical optimization improvements, circuit and layout optimizations, and error mitigation protocols.
Specifically, we employed an effective Hamiltonian to reduce the number of qubits, optimized the variational ansatz to yield shallower quantum circuits, enhanced classical optimization to better tolerate stochastic fluctuations in energy estimations, and implemented automatic selection of optimal layouts and CNOT gates. 
Measurement errors were mitigated using techniques natively supported by Qiskit, including zero-noise extrapolation and readout error correction.

Despite these optimizations, our results clearly show that for quantum chemistry Hamiltonians of medium-sized molecules, exemplified by the benzene molecule, the accuracy achievable on current hardware remains well below the threshold of chemical accuracy (1 kcal/mol). 
The intrinsic noise in energy evaluations obtained from quantum hardware prevents reliable convergence of full variational optimizations for such molecular systems.
Through simulations incorporating realistic noise models, we have identified insufficient qubit coherence times as a major source of error. Our estimations indicate that current quantum hardware coherence times must be improved by at least two orders of magnitude for VQE simulations to achieve the required accuracy for reliable and predictive quantum chemistry calculation. This underscores a critical hardware threshold that needs to be surpassed for VQEs to become truly transformative tools in molecular science.

\section{Methods}\label{sec:methods}

The limitations of current quantum hardware make it extremely challenging
to perform a full VQE optimization of medium-sized molecules directly on a quantum processor. 
For this reason, in this work, we focused on evaluating the energy accuracy of approximate wave functions obtained at each
iteration of the ADAPT-VQE algorithm. 
These approximate wave functions were extracted from an ideal (exact)
Qubit-ADAPT simulation. As a reference, we employed the natural orbitals 
from an unrestricted Hartree–Fock (UHF) calculation. To reduce the 
quantum resource requirements, we defined an active space of four orbitals and four electrons for the benzene molecule, resulting in quantum circuits acting on 8 qubits.

The geometry of the benzene molecule was retrieved from the PubChem 
database.\cite{pubchem} Molecular orbitals used to construct the Hamiltonian were  computed with the \texttt{PySCF} package \cite{pyscf} using a STO-3G basis set. The  mapping from molecular orbitals to qubits
is performed using the Jordan–Wigner transformation. \cite{Jordan1928} The
operator pool employed in the ADAPT-VQE procedure consists of unitary coupled-cluster (UCC) operators restricted to occupied-to-virtual spin-singlet adapted single and double excitations.\cite{Grimsley2019}
Simulations were carried out with an in-house Python implementation of ADAPT-VQE, relying on the \texttt{NumPy},\cite{numpy}
\texttt{SciPy},\cite{2020SciPy} and 
\texttt{OpenFermion}\cite{mcclean2019openfermion} libraries.

Calculations on real quantum hardware were performed on the \texttt{IBM\_tornino} device. Circuit implementation and transpilation were done using \texttt{Qiskit}, with circuit optimization level 3. Qubit layout was optimized by selecting the best available correlative physical qubits for each run. Measurements were executed with built-in error mitigation techniques provided by IBM's runtime using resilience level 2. These include Twirled Readout Error eXtinction (TREX), measurement twirling, Zero Noise Extrapolation (ZNE), and gate twirling.

\section*{Supplementary information} 

Further details on: effective Hamiltonian, compression of electronic Hamiltonians, comparison between Fermion and Qubit operator pools, classical optimizer, circuit optimization, quantum hardware experiments, and instruction execution times.

\section*{Acknowledgments}

This work was funded by the MICIU/AEI/10.13039/501100011033 (project PID2022-136231NB-I00) and by FEDER, UE, the European Union (project NextGenerationEU/PRTR-C17.I1), as well as by the IKUR Strategy under the collaboration agreement between Ikerbasque Foundation and DIPC on behalf of the Department of Education of the Basque Government.
The authors acknowledge the financial support received from the BasQ Strategy under the collaboration agreement between Ikerbasque Foundation and DIPC on behalf of the Department of Education of the Basque Government.
The authors are thankful for the technical and human support provided by the DIPC Computer Center. 
D.C. is thankful for financial support from IKERBASQUE (Basque Foundation for Science).

\section*{Ethics declarations}

The authors declare no competing interests.


\clearpage
\bibliography{references}

\begin{thebibliography}{10}
\expandafter\ifx\csname url\endcsname\relax
  \def\url#1{\burl{#1}}\fi
\expandafter\ifx\csname urlprefix\endcsname\relax\def\urlprefix{URL }\fi
\providecommand{\bibinfo}[2]{#2}
\providecommand{\eprint}[2][]{\url{#2}}
\providecommand{\doi}[1]{\url{https://doi.org/#1}}
\bibcommenthead

\bibitem{Feynman:1982}
\bibinfo{author}{Feynman, R.~P.}
\newblock \bibinfo{title}{Simulating physics with computers} \textbf{\bibinfo{volume}{21}}, \bibinfo{pages}{467--488} (\bibinfo{year}{1982}).
\newblock \urlprefix\url{https://doi.org/10.1007/BF02650179}.

\bibitem{Aspuru:2005}
\bibinfo{author}{Aspuru-Guzik, A.}, \bibinfo{author}{Dutoi, A.~D.}, \bibinfo{author}{Love, P.~J.} \& \bibinfo{author}{Head-Gordon, M.}
\newblock \bibinfo{title}{Simulated quantum computation of molecular energies} \textbf{\bibinfo{volume}{309}}, \bibinfo{pages}{1704--1707} (\bibinfo{year}{2005}).

\bibitem{Cao:review:2019}
\bibinfo{author}{Cao, Y.} \emph{et~al.}
\newblock \bibinfo{title}{Quantum chemistry in the age of quantum computing} \textbf{\bibinfo{volume}{119}}, \bibinfo{pages}{10856--10915} (\bibinfo{year}{2019}).

\bibitem{ibm:whitepaper:2024}
\bibinfo{author}{Alexeev, Y.} \emph{et~al.}
\newblock \bibinfo{title}{Quantum-centric supercomputing for materials science: A perspective on challenges and future directions}.
\newblock \emph{\bibinfo{journal}{Future Gener. Comput. Syst.}} \textbf{\bibinfo{volume}{160}}, \bibinfo{pages}{666--710} (\bibinfo{year}{2024}).
\newblock \urlprefix\url{https://www.sciencedirect.com/science/article/pii/S0167739X24002012}.

\bibitem{Peruzzo:2014}
\bibinfo{author}{Peruzzo, A.} \emph{et~al.}
\newblock \bibinfo{title}{A variational eigenvalue solver on a photonic quantum processor} \textbf{\bibinfo{volume}{5}}, \bibinfo{pages}{4213} (\bibinfo{year}{2014}).

\bibitem{vqe_review:2022}
\bibinfo{author}{Tilly, J.} \emph{et~al.}
\newblock \bibinfo{title}{The variational quantum eigensolver: A review of methods and best practices} \textbf{\bibinfo{volume}{986}}, \bibinfo{pages}{1--128} (\bibinfo{year}{2022}).
\newblock \bibinfo{note}{The Variational Quantum Eigensolver: a review of methods and best practices}.

\bibitem{Fedorov:vqe_review:2022}
\bibinfo{author}{Fedorov, D.~A.}, \bibinfo{author}{Peng, B.}, \bibinfo{author}{Govind, N.} \& \bibinfo{author}{Alexeev, Y.}
\newblock \bibinfo{title}{Vqe method: a short survey and recent developments} \textbf{\bibinfo{volume}{6}}, \bibinfo{pages}{2} (\bibinfo{year}{2022}).

\bibitem{Preskill2018quantumcomputingi}
\bibinfo{author}{Preskill, J.}
\newblock \bibinfo{title}{Quantum {C}omputing in the {NISQ} era and beyond}.
\newblock \emph{\bibinfo{journal}{{Quantum}}} \textbf{\bibinfo{volume}{2}}, \bibinfo{pages}{79} (\bibinfo{year}{2018}).
\newblock \urlprefix\url{https://doi.org/10.22331/q-2018-08-06-79}.

\bibitem{Grimsley2019}
\bibinfo{author}{Grimsley, H.~R.}, \bibinfo{author}{Economou, S.~E.}, \bibinfo{author}{Barnes, E.} \& \bibinfo{author}{Mayhall, N.~J.}
\newblock \bibinfo{title}{{An adaptive variational algorithm for exact molecular simulations on a quantum computer}} \textbf{\bibinfo{volume}{10}}, \bibinfo{pages}{3007} (\bibinfo{year}{2019}).
\newblock \urlprefix\url{https://doi.org/10.1038/s41467-019-10988-2 https://www.nature.com/articles/s41467-019-10988-2}.

\bibitem{Blum_2024}
\bibinfo{author}{Blum, V.} \emph{et~al.}
\newblock \bibinfo{title}{Roadmap on methods and software for electronic structure based simulations in chemistry and materials} \textbf{\bibinfo{volume}{6}}, \bibinfo{pages}{042501} (\bibinfo{year}{2024}).

\bibitem{integral_thresh}
\bibinfo{note}{It is worth noting that an energy threshold of $10^{-2}$ Hartrees for individual Hamiltonian terms corresponds to even smaller overall contributions to the ground-state energy, as the one- and two-electron terms are scaled by the one- and two-particle density matrices, respectively.}

\bibitem{Vaquero:VQEs:2024}
\bibinfo{author}{Vaquero-Sabater, N.}, \bibinfo{author}{Carreras, A.}, \bibinfo{author}{Or\'us, R.}, \bibinfo{author}{Mayhall, N.~J.} \& \bibinfo{author}{Casanova, D.}
\newblock \bibinfo{title}{Physically motivated improvements of variational quantum eigensolvers} \textbf{\bibinfo{volume}{20}}, \bibinfo{pages}{5133--5144} (\bibinfo{year}{2024}).
\newblock \urlprefix\url{https://doi.org/10.1021/acs.jctc.4c00329}.

\bibitem{Grimsley:adapt:2023}
\bibinfo{author}{Grimsley, H.~R.}, \bibinfo{author}{Barron, G.~S.}, \bibinfo{author}{Barnes, E.}, \bibinfo{author}{Economou, S.~E.} \& \bibinfo{author}{Mayhall, N.~J.}
\newblock \bibinfo{title}{{Adaptive, problem-tailored variational quantum eigensolver mitigates rough parameter landscapes and barren plateaus}} \textbf{\bibinfo{volume}{9}}, \bibinfo{pages}{19} (\bibinfo{year}{2023}).

\bibitem{Tang2021}
\bibinfo{author}{Tang, H.~L.} \emph{et~al.}
\newblock \bibinfo{title}{{Qubit-ADAPT-VQE: An Adaptive Algorithm for Constructing Hardware-Efficient Ans{\"{a}}tze on a Quantum Processor}} \textbf{\bibinfo{volume}{2}}, \bibinfo{pages}{020310} (\bibinfo{year}{2021}).
\newblock \urlprefix\url{https://link.aps.org/doi/10.1103/PRXQuantum.2.020310}.

\bibitem{claudino_benchmarking_2020}
\bibinfo{author}{Claudino, D.}, \bibinfo{author}{Wright, J.}, \bibinfo{author}{{McCaskey}, A.~J.} \& \bibinfo{author}{Humble, T.~S.}
\newblock \bibinfo{title}{Benchmarking adaptive variational quantum eigensolvers} \textbf{\bibinfo{volume}{8}}, \bibinfo{pages}{606863}.
\newblock \urlprefix\url{https://www.frontiersin.org/articles/10.3389/fchem.2020.606863/full}.

\bibitem{singh_benchmarking_2023}
\bibinfo{author}{Singh, H.}, \bibinfo{author}{Majumder, S.} \& \bibinfo{author}{Mishra, S.}
\newblock \bibinfo{title}{Benchmarking of different optimizers in the variational quantum algorithms for applications in quantum chemistry}.
\newblock \emph{\bibinfo{journal}{J. Comp. Phys.}} \textbf{\bibinfo{volume}{159}}, \bibinfo{pages}{044117}.

\bibitem{Powell:cobyla:1994}
\bibinfo{author}{Powell, M. J.~D.}
\newblock \emph{\bibinfo{title}{A Direct Search Optimization Method That Models the Objective and Constraint Functions by Linear Interpolation}}, \bibinfo{pages}{51--67} (\bibinfo{publisher}{Springer Netherlands}, \bibinfo{address}{Dordrecht}, \bibinfo{year}{1994}).

\bibitem{McClean_2016}
\bibinfo{author}{McClean, J.~R.}, \bibinfo{author}{Romero, J.}, \bibinfo{author}{Babbush, R.} \& \bibinfo{author}{Aspuru-Guzik, A.}
\newblock \bibinfo{title}{The theory of variational hybrid quantum-classical algorithms} \textbf{\bibinfo{volume}{18}}, \bibinfo{pages}{023023} (\bibinfo{year}{2016}).

\bibitem{mansky_decomposition_2023}
\bibinfo{author}{Mansky, M.~B.}, \bibinfo{author}{Puigvert, V.~R.}, \bibinfo{author}{Castillo, S.~L.} \& \bibinfo{author}{Linnhoff-Popien, C.}
\newblock \bibinfo{title}{Decomposition algorithm of an arbitrary pauli exponential through a quantum circuit}.
\newblock \eprint{2305.04807 [quant-ph]}.

\bibitem{GiurgicaTiron2020}
\bibinfo{author}{Giurgica-Tiron, T.}, \bibinfo{author}{Hindy, Y.}, \bibinfo{author}{LaRose, R.}, \bibinfo{author}{Mari, A.} \& \bibinfo{author}{Cincio, L.}
\newblock \bibinfo{title}{Digital zero noise extrapolation for quantum error mitigation} \textbf{\bibinfo{volume}{2}}, \bibinfo{pages}{010317} (\bibinfo{year}{2021}).

\bibitem{qiskit_noise_models}
\bibinfo{title}{{IBM} {Q}uantum {D}ocumentation: Build noise models}.
\newblock \urlprefix\url{https://docs.quantum.ibm.com/guides/build-noise-models}.

\bibitem{pubchem}
\bibinfo{author}{Kim, S.} \emph{et~al.}
\newblock \bibinfo{title}{Pubchem 2025 update}.
\newblock \emph{\bibinfo{journal}{Nucleic Acids Research}} \textbf{\bibinfo{volume}{53}}, \bibinfo{pages}{D1516--D1525} (\bibinfo{year}{2024}).
\newblock \urlprefix\url{https://doi.org/10.1093/nar/gkae1059}.

\bibitem{pyscf}
\bibinfo{author}{Sun, Q.} \emph{et~al.}
\newblock \bibinfo{title}{Pyscf: the python-based simulations of chemistry framework} \textbf{\bibinfo{volume}{8}}, \bibinfo{pages}{e1340} (\bibinfo{year}{2018}).
\newblock \urlprefix\url{https://wires.onlinelibrary.wiley.com/doi/abs/10.1002/wcms.1340}.

\bibitem{Jordan1928}
\bibinfo{author}{Jordan, P.} \& \bibinfo{author}{Wigner, E.}
\newblock \bibinfo{title}{{\"U}ber das paulische {\"a}quivalenzverbot}.
\newblock \emph{\bibinfo{journal}{Zeitschrift f{\"u}r Physik}} \textbf{\bibinfo{volume}{47}}, \bibinfo{pages}{631--651} (\bibinfo{year}{1928}).
\newblock \urlprefix\url{https://doi.org/10.1007/BF01331938}.

\bibitem{numpy}
\bibinfo{author}{Harris, C.~R.} \emph{et~al.}
\newblock \bibinfo{title}{Array programming with {NumPy}}.
\newblock \emph{\bibinfo{journal}{Nature}} \textbf{\bibinfo{volume}{585}}, \bibinfo{pages}{357--362} (\bibinfo{year}{2020}).
\newblock \urlprefix\url{https://doi.org/10.1038/s41586-020-2649-2}.

\bibitem{2020SciPy}
\bibinfo{author}{Virtanen, P.} \emph{et~al.}
\newblock \bibinfo{title}{{SciPy} 1.0: Fundamental algorithms for scientific computing in python}.
\newblock \emph{\bibinfo{journal}{Nature Methods}} \textbf{\bibinfo{volume}{17}}, \bibinfo{pages}{261--272} (\bibinfo{year}{2020}).

\bibitem{mcclean2019openfermion}
\bibinfo{author}{McClean, J.~R.} \emph{et~al.}
\newblock \bibinfo{title}{Openfermion: The electronic structure package for quantum computers} (\bibinfo{year}{2019}).
\newblock \eprint{1710.07629}.

\end{thebibliography}

\end{document}


\tableofcontents

\clearpage

\section{Effective Hamiltonian} \label{sec:SI_Heff}

To construct the effective Hamiltonian, we apply the frozen core approximation. In this approach, the molecular spin-orbitals are divided into three sets:

\noindent-- \textbf{Core spin-orbitals}:  low-energy spin-orbitals that are treated as an effective potential in the Hamiltonian; 

\noindent-- \textbf{Active spin-orbitals}: which are treated explicitly in the Hamiltonian;  

\noindent--  \textbf{Truncated spin-orbitals}: corresponding to high-energy spin-orbitals that are not included in the Hamiltonian.

In this work we used a spin-orbital basis, where spin-up (\(\alpha\)) and spin-down (\(\beta\)) components are interleaved. In this basis the effective Hamiltonian is defined as

\begin{equation} 
   \hat{\mathcal{H}}_\text{eff} = \sum_{pq}^\text{act}{\tilde h_{pq} \hat a_p^\dagger \hat a_q} + \frac{1}{2}\sum_{pqrs}^\text{act}{g_{pqrs} \hat a_p^\dagger \hat a_q^\dagger \hat a_r \hat a_s} + \hat{V}_\text{eff}, 
   \label{eq:h_eff}
\end{equation}

where \( \tilde h_{pq} \) are the effective one-electron integrals, defined as

\begin{equation}
     \tilde h_{pq} = h_{pq} + \sum_{i}^{\text{core}} \left( g_{iqpi} - g_{iipq} \right),
\end{equation}

with \( h_{pq} \) denoting the one-electron integrals from the original Hamiltonian; and the effective potential term \( \hat{V}_\text{eff} \), accounting for the core spin-orbitals, is given by

\begin{equation}
     \hat{V}_\text{eff} = \sum_{i}^{\text{core}} h_{ii} + \frac{1}{2}\sum_{ij}^{\text{core}} \left( g_{jiij} - g_{jjii} \right).
\end{equation}

Note that indices \(i,j\) run over the core spin-orbitals, while \(p,q,r,s\) run over the active spin-orbitals. 

\section{Compression of electronic Hamiltonians} \label{sec:SI_compression}
\begin{figure}[H]
\includegraphics[width=6.5cm]{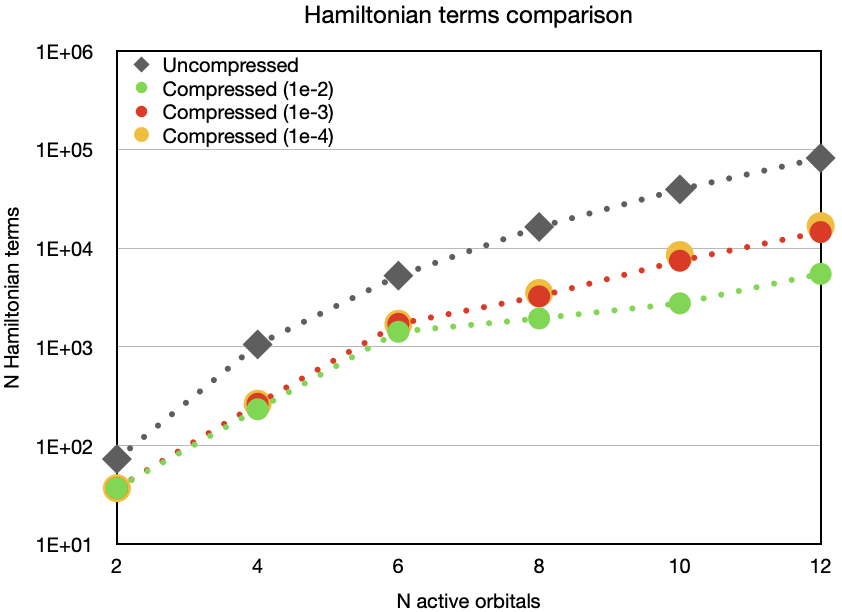}
\includegraphics[width=6.5cm]{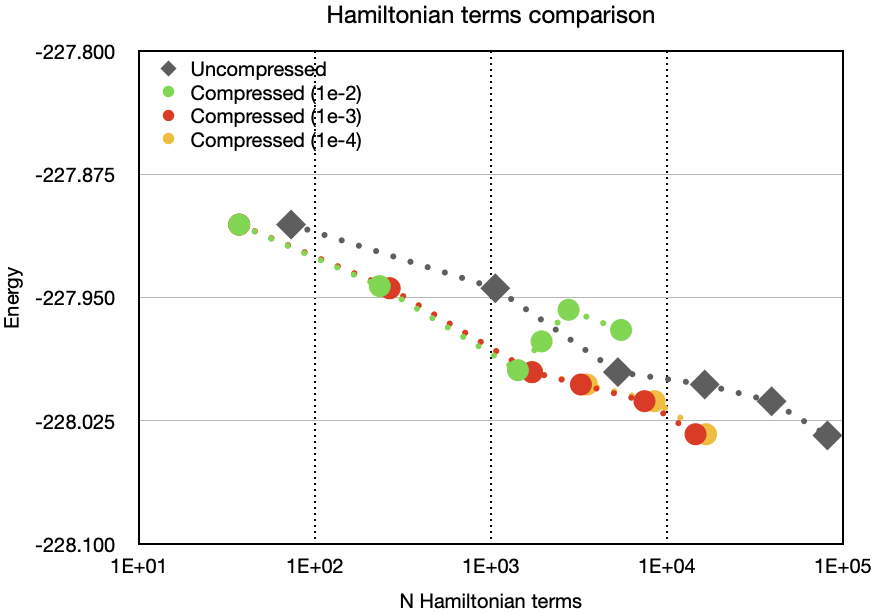}
  \caption{Analysis of the impact of Hamiltonian compression on the number of Hamiltonian terms and the ground state energy. On the left, we show the number of Hamiltonian terms as a function of the active space size for different Hamiltonian with three compression thresholds (1e-2, 1e-3, and 1e-4), compared to the uncompressed Hamiltonian. On the right, we plot the ground state (GS) energy as a function of the number of Hamiltonian terms. The GS energies have been computed using the DMRG method (with a bond dimension of 200) for three different compression thresholds (1e-3, 1e-4, and 1e-5). The Hamiltonian is expressed in the basis of natural Orbitals.}
  \label{fig:compression_analysis}
\end{figure}

\begin{figure}[H]
\includegraphics[width=6.5cm]{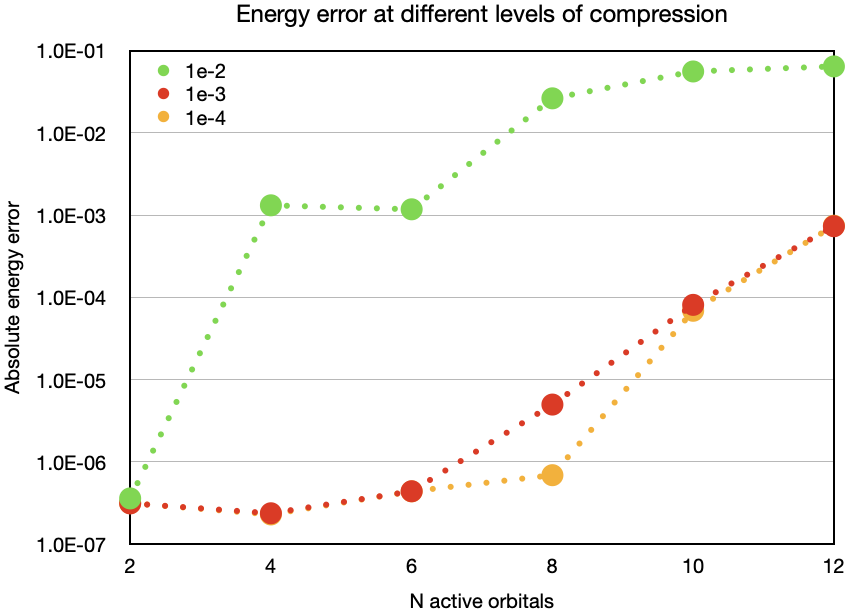}
\includegraphics[width=6.5cm]{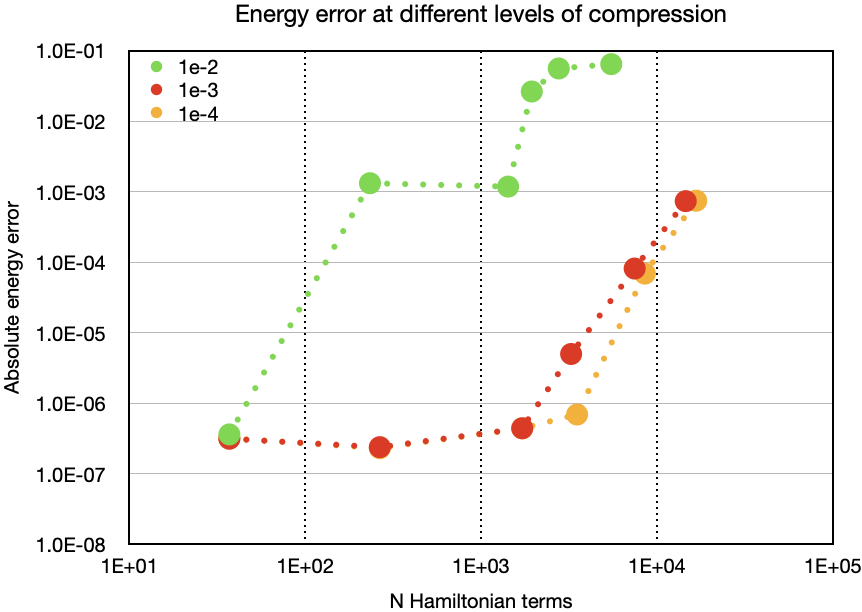}
  \caption{ Analysis of the energy error with respect to different Hamiltonian compression thresholds (1e-2, 1e-3, and 1e-4). On the left, we show the ground state energy error as a function of the active space size. On the right, we present the GS energy error as a function of the number of Hamiltonian terms. The ground state energy energies have been computed using the DMRG method with a bond dimension of 200. The Hamiltonian is expressed in the basis of natural Orbitals.}
  \label{fig:compression_analysis_error}
\end{figure}

\begin{figure}[H]
\includegraphics[width=6.5cm]{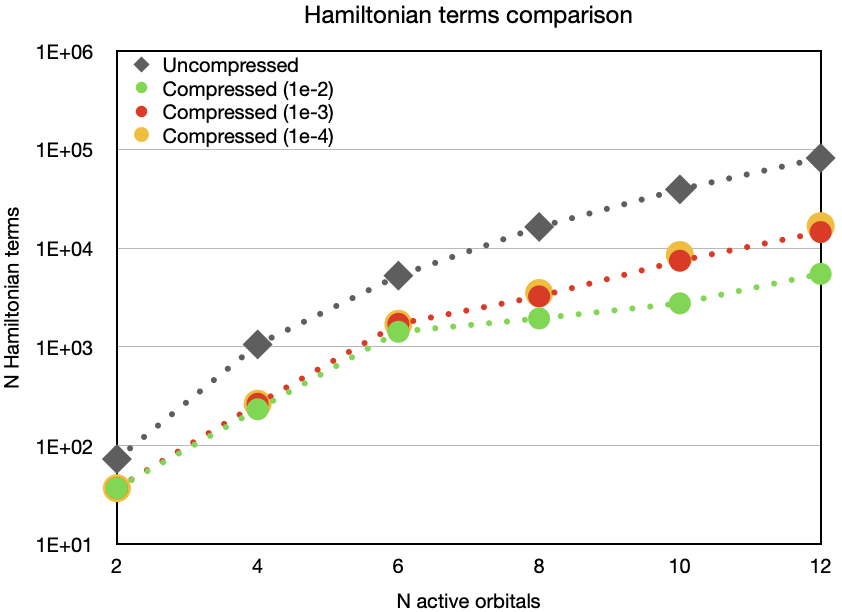}
\includegraphics[width=6.5cm]{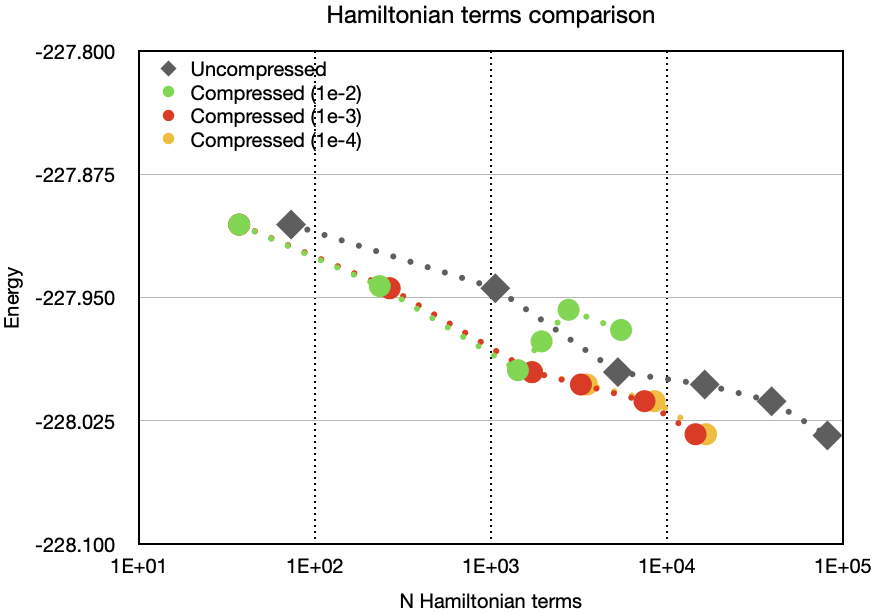}
  \caption{Analysis of the impact of Hamiltonian compression on the number of Hamiltonian terms and the ground state energy. On the left, we show the number of Hamiltonian terms as a function of the active space size for different Hamiltonian with three compression thresholds (1e-2, 1e-3, and 1e-4), compared to the uncompressed Hamiltonian. On the right, we plot the ground state (GS) energy as a function of the number of Hamiltonian terms. The GS energies have been computed using the DMRG method (with a bond dimension of 200) for three different compression thresholds (1e-3, 1e-4, and 1e-5). The Hamiltonian is expressed in the basis of canonical orbitals.}
  \label{fig:compression_analysis_can}
\end{figure}

In this section, we analyze the impact of Hamiltonian compression on the ground state energy. Additionally, we examine how the size of the Hamiltonian evolves as the number of active space orbitals increases, considering different compression thresholds. For this study, we construct the Hamiltonian for increasing active space sizes, ranging from 2 to 12 molecular orbitals, and apply different compression thresholds (1e-2, 1e-3, and 1e-4). The results are calculated using canonical orbitals (Figures \ref{fig:compression_analysis_can} and \ref{fig:compression_analysis_can}), and using natural orbitals \ref{fig:compression_analysis} and \ref{fig:compression_analysis_error}.

Focusing in natural orbitals, figure \ref{fig:compression_analysis} (left) shows the number of Hamiltonian terms as a function of the active space size. As expected, the number of terms increases with the active space size. For small active spaces, there is almost no difference between the three compression thresholds. However, beyond 8 orbitals, a significant difference emerges between 1e-2 and the other two (1e-3 and 1e-4). In this regime, the number of terms is notably reduced for 1e-2, while for 1e-3 and 1e-4, the difference remains marginal.

Figure \ref{fig:compression_analysis} (right) illustrates the impact of compression from an energy perspective by plotting the ground state (GS) energy as a function of the number of Hamiltonian terms. The results indicate that a compressed Hamiltonian provides a better GS energy-to-terms ratio compared to the uncompressed Hamiltonian. This trend holds for all compression thresholds except 1e-2, where, around 1000 terms, the ratio deteriorates, leading to a GS energy higher than that of the uncompressed Hamiltonian for the same number of terms.

In Figure \ref{fig:compression_analysis_error}, we analyze the evolution of the GS energy error (defined as the difference between the GS energy of the compressed and uncompressed Hamiltonians) with respect to both the active space size (left) and the number of Hamiltonian terms (right). Even for a compression threshold of 1e-2, the GS energy error remains within chemical accuracy up to an active space of 6 orbitals. For 1e-3 compression, the GS energy remains below chemical accuracy up to 12 molecular orbitals. Notably, the difference between 1e-3 and 1e-4 is minimal.

Examining the energy error as a function of the number of Hamiltonian terms, we observe that a 1e-3 compression is highly efficient up to 6 active orbitals, yielding a Hamiltonian with negligible loss relative to the uncompressed one. Beyond this point, the energy error increases exponentially with both the number of active orbitals and the number of Hamiltonian terms.

These results suggest that as the active space size increases—and thus the accuracy of the GS energy improves—small terms become more relevant, limiting the extent to which compression can be applied. However, for active spaces of up to 6 orbitals, a compression threshold of 1e-2 appears sufficient to maintain the GS energy error below chemical accuracy.

\begin{figure}[H]
\includegraphics[width=6.5cm]{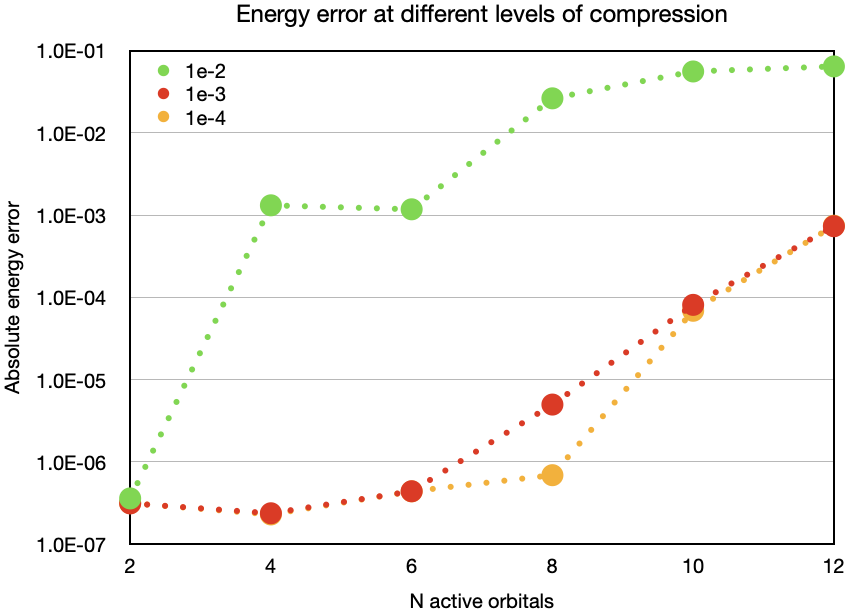}
\includegraphics[width=6.5cm]{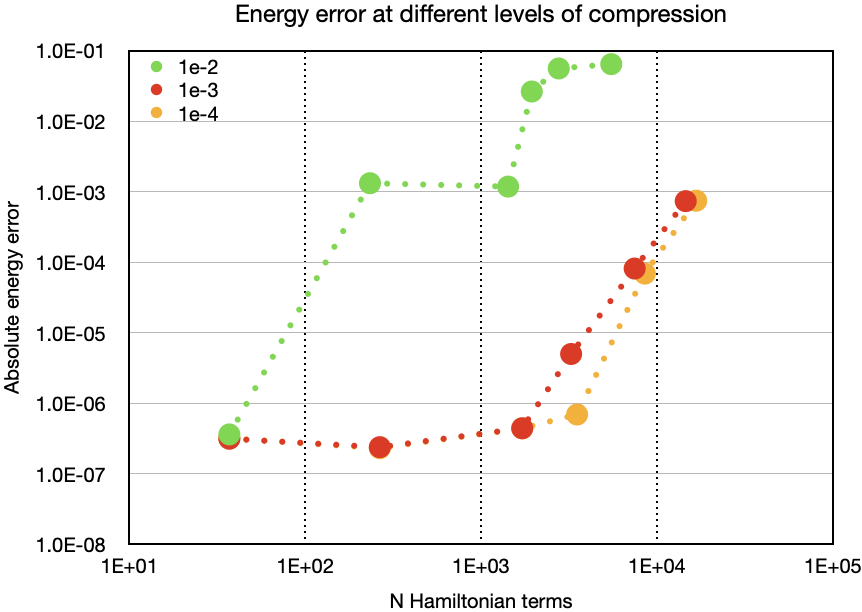}
  \caption{ Analysis of the energy error with respect to different Hamiltonian compression thresholds (1e-2, 1e-3, and 1e-4). On the left, we show the ground state energy error as a function of the active space size. On the right, we present the GS energy error as a function of the number of Hamiltonian terms. The ground state energy energies have been computed using the DMRG method with a bond dimension of 200. The Hamiltonian is expressed in the basis of canonical Orbitals.}
  \label{fig:compression_analysis_error_can}
\end{figure}

Examining the results obtained using canonical orbitals (Figures \ref{fig:compression_analysis_can}  and \ref{fig:compression_analysis_error_can}), we observe that the use of canonical orbitals is detrimental compared to the use of natural orbitals. The energy-to-number-of-terms ratio presented  in Figure \ref{fig:compression_analysis_can} (right) is significantly worse than that observed with  natural orbitals, where compression provides little to no benefit at any level. Furthermore, we  observe that the GS energy error with respect to the uncompressed Hamiltonian (Figure \ref{fig:compression_analysis_error_can}) is substantially higher when using canonical orbitals, frequently exceeding chemical accuracy.  

This suggests that the use of natural orbitals leads to Hamiltonians where the most important interactions are concentrated in a smaller number of terms, making the compression procedure more effective. 
However, it is important to note that as higher precision is required, the contribution of small Hamiltonian terms becomes more significant, and their impact  on the ground state energy may no longer be negligible.  

\section{Fermion vs. Qubit operator pools} \label{sec:SI_oper_pools}

\begin{figure}[H]
\includegraphics[width=8cm]{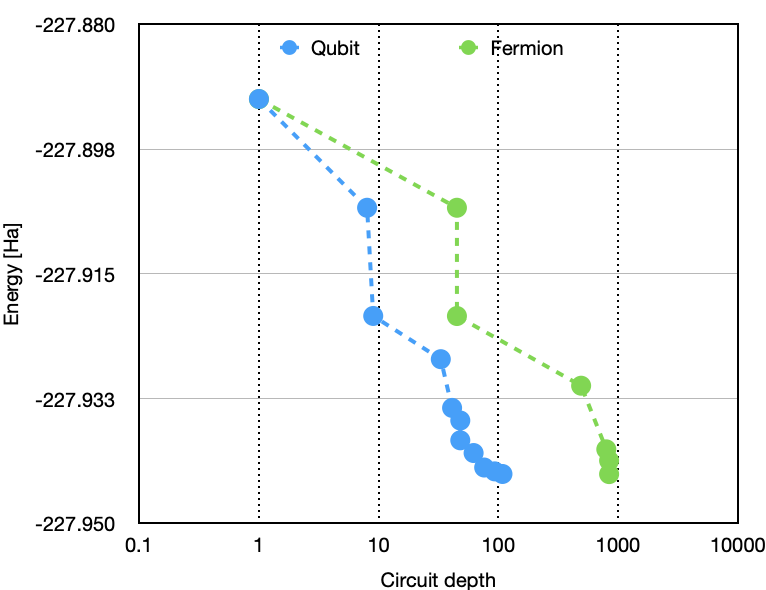}
  \caption{Ground state energy (in Hartrees) as a function of the number of iterations for the simulation of the benzene molecule with fermion (green) and qubit (blue) ADAPT-VQE algorithm.}
  \label{si:fig:vqe_comparison}
\end{figure}

\begin{figure}[H]
\includegraphics[width=7.5cm]{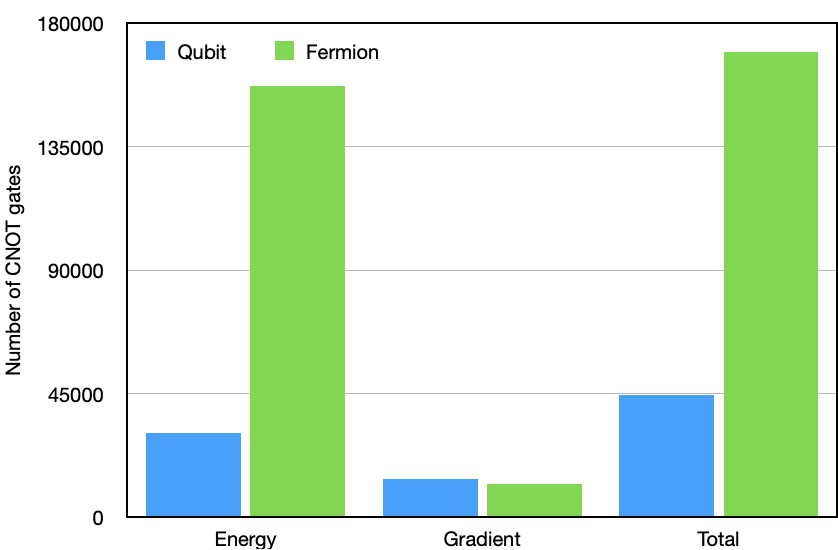}
\includegraphics[width=7.5cm]{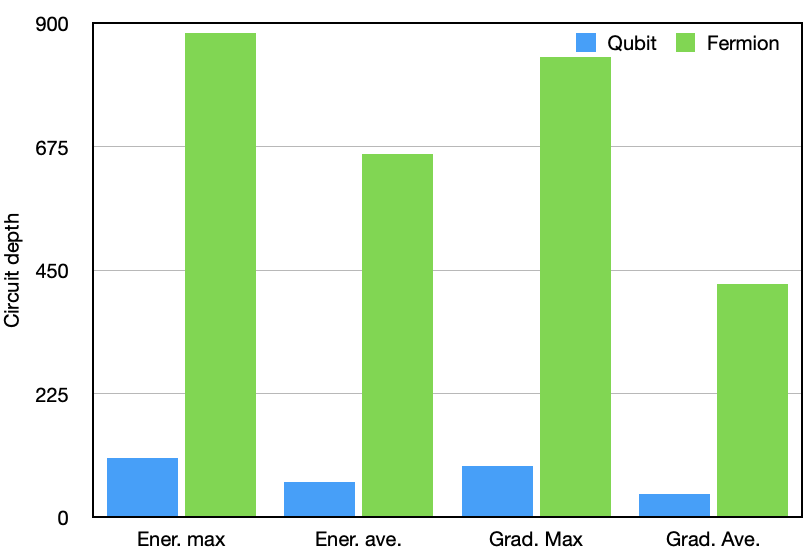}
  \caption{Comparison of the computational cost of VQE algorithms using fermion vs qubit operators for benzene molecule. On the left there is a break down of the total number of CNOTS used for all circuits in the full simulation. We separate this number in energy evaluations and gradient evaluations. On the right there is an analysis of the circuit depths of the circuits used to evaulate the energy and the gradients. We show the maxium circuit depth and the average circuit depth. }
  \label{si:fig:vqe_circuit}
\end{figure}

\begin{figure}[H]
\includegraphics[width=7.5cm]{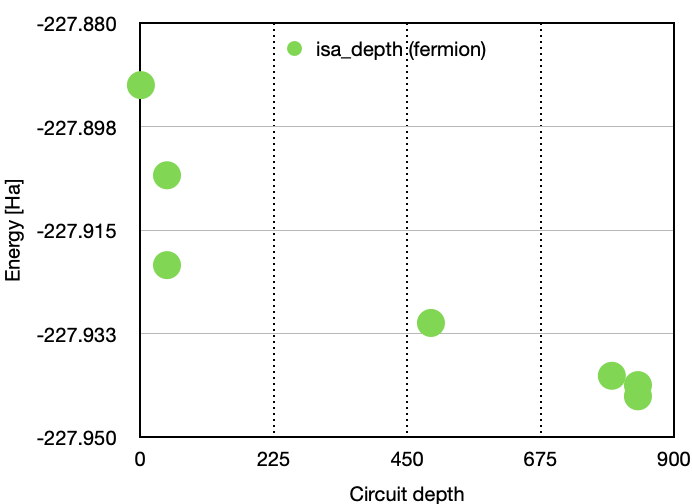}
\includegraphics[width=7.5cm]{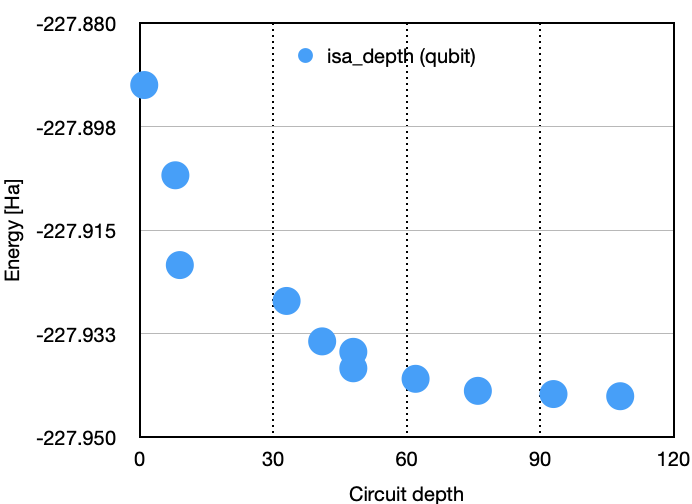}
  \caption{Comparison of the circuit depth (of the circuit used to evaluate the energy) vs energy obtained from an ADAPT-VQE simulation (exact). On the left  fermion-ADAPT-VQE, on the right qubit-ADAPT-VQE.}
  \label{fig:vqe_circuit_depth_ener}
\end{figure}

\section{Classical optimizer} \label{sec:SI_class_opt}

\begin{figure}[H]
\includegraphics[width=6.5cm]{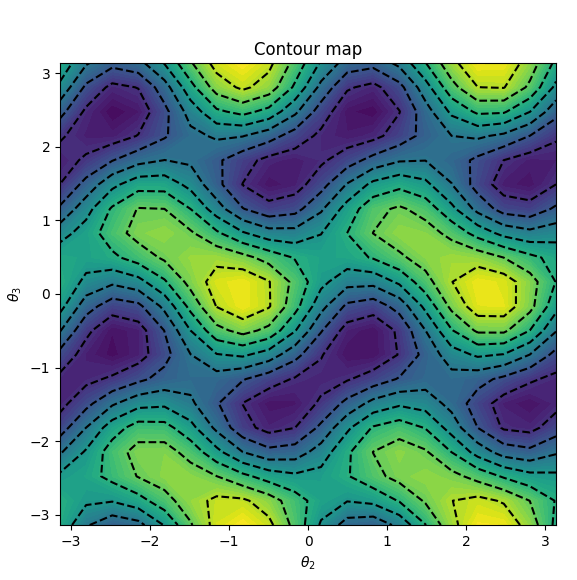}
\includegraphics[width=6.5cm]{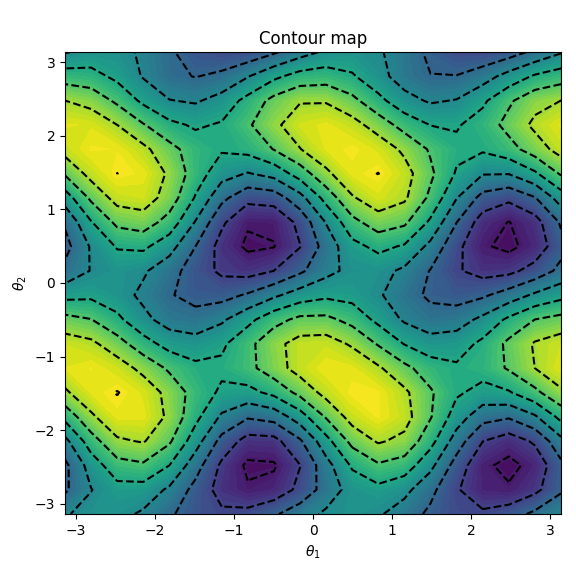}
  \caption{Energy surface generated by the scan over the coefficients of a two operator qubit-ADAPT-VQE ansatz of \ce{H4} molecule. Left figure shows the scan along coefficients 1 and 2 and right figure is the scan along coefficients 2 and 3}
  \label{si:fig:operators_map_h4}
\end{figure}

\begin{figure}[H]
  \includegraphics[width=8cm]{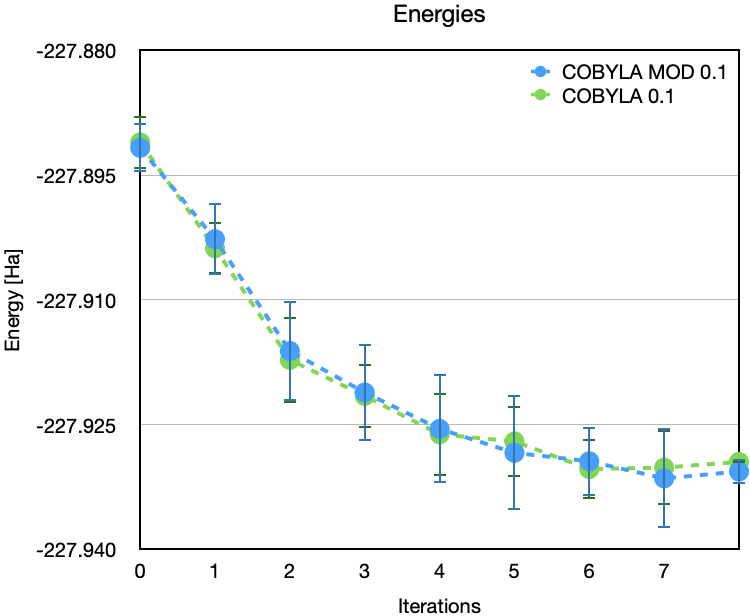}
  \includegraphics[width=8cm]{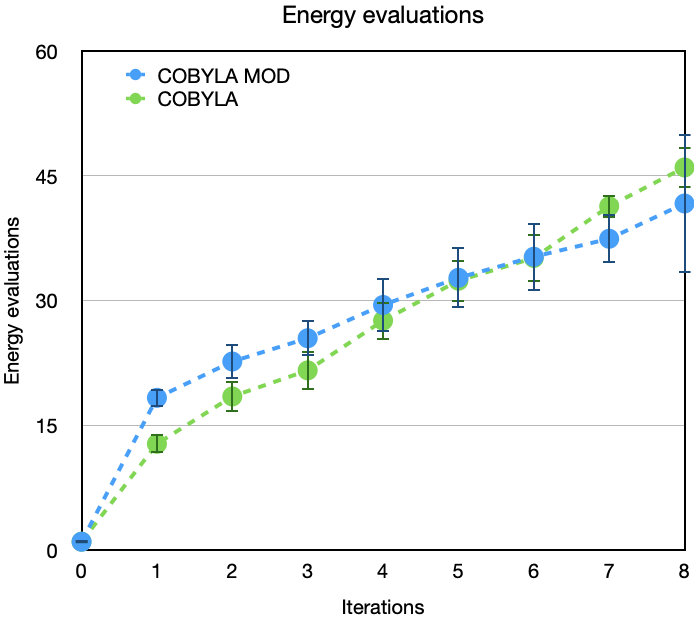}
  \caption{(a) Ground state energy and (b) energy evaluations of benzene molecule along the qubit-ADAPT iterations using COBYLA (green) and COBYLA-MOD (blue). Each point corresponds to the average of 30 simulations using 1000 shots. Default tolerance in standard COBYLA is 10$^{-3}$ Hartrees.}
  \label{si:fig:opt_1000}
\end{figure}

\begin{figure}[H]
 \includegraphics[width=8cm]{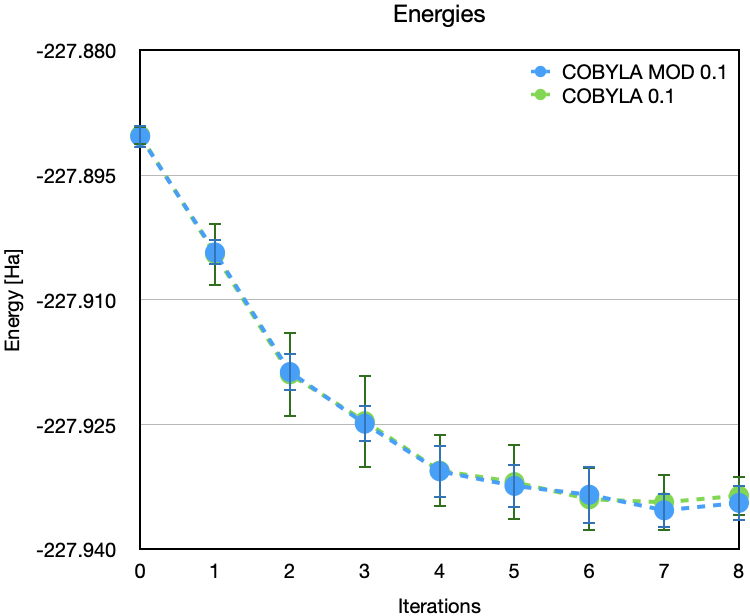}
 \includegraphics[width=8cm]{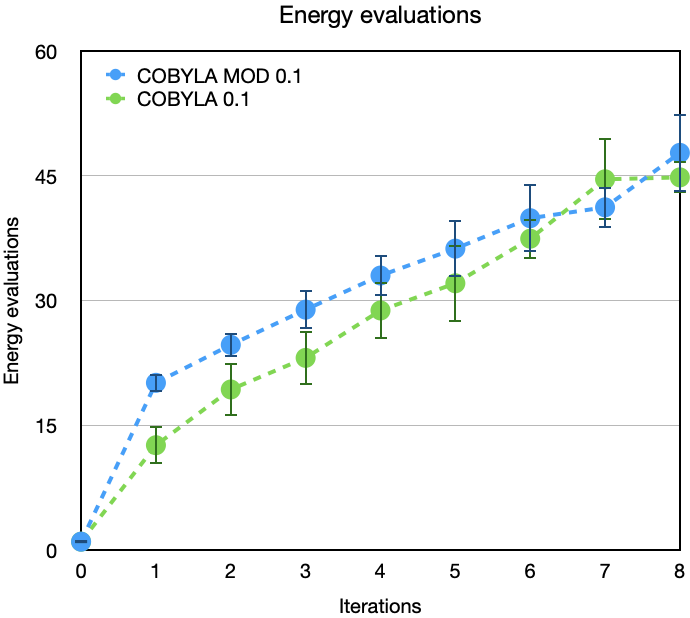}
  \caption{(a) Ground state energy and (b) energy evaluations of benzene molecule along the qubit-ADAPT iterations using COBYLA (green) and COBYLA-MOD (blue). Each point corresponds to the average of 30 simulations using 5000 shots. Default tolerance in standard COBYLA is 10$^{-3}$ Hartrees.}
  \label{si:fig:opt_5000}
\end{figure}

\subsection{Stretched \ce{H4} model} \label{sec:cobyla_H4}

\begin{figure}[H]
 \includegraphics[width=8cm]{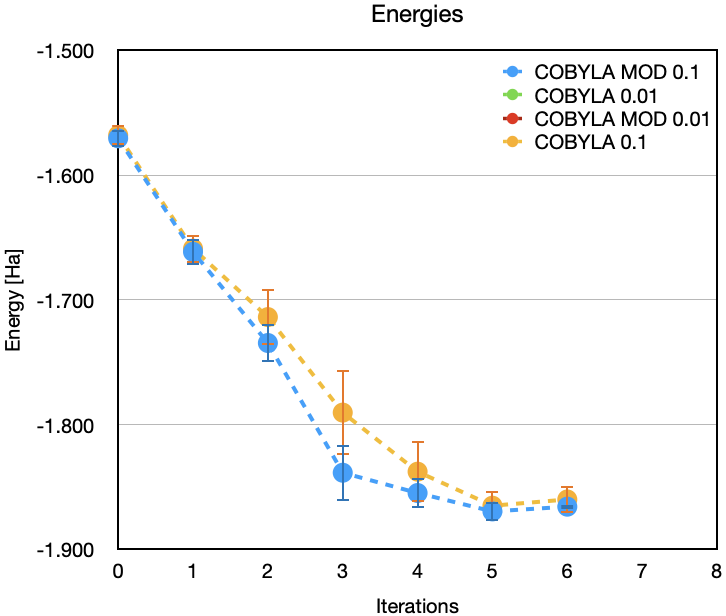}
 \includegraphics[width=8cm]{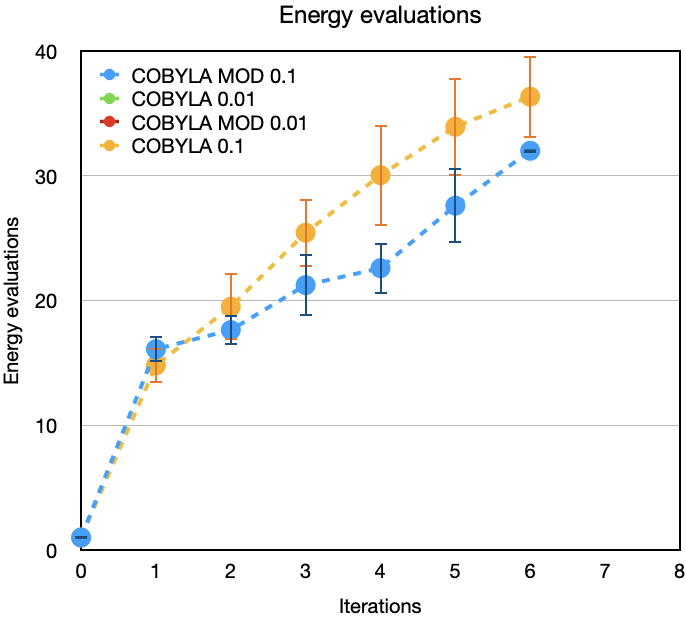}
  \caption{Molecule \ce{H4}., Comparison of the optimization using COBYLA and COBYLA-MOD using a value of 0.1 for rhobeg. Number of shots is 1000. Default tolerance is 1e-3 Ha.}
  \label{si:fig:opt_h4_1000}
\end{figure}

To test the COBLYLA-MOD with highly correlated system we have analyzed the linear H4 molecule with bond distance of 2\AA. For this example we have also used Qiskit Aer simulator to compute the energy with 1000 shots. In figure \ref{si:fig:opt_h4_1000} we show a comparison of a  the average of 30 qubit-ADAPT-VQE simulations using both COBYLA and COBYLA-MOD. In the left plot we show the energy as a function of each iteration step and on the right we plot the number of energy evaluations. For reference we include two exact simulation computed using Fermion operators and qubit operators.

\begin{figure}[H]
 \includegraphics[width=8cm]{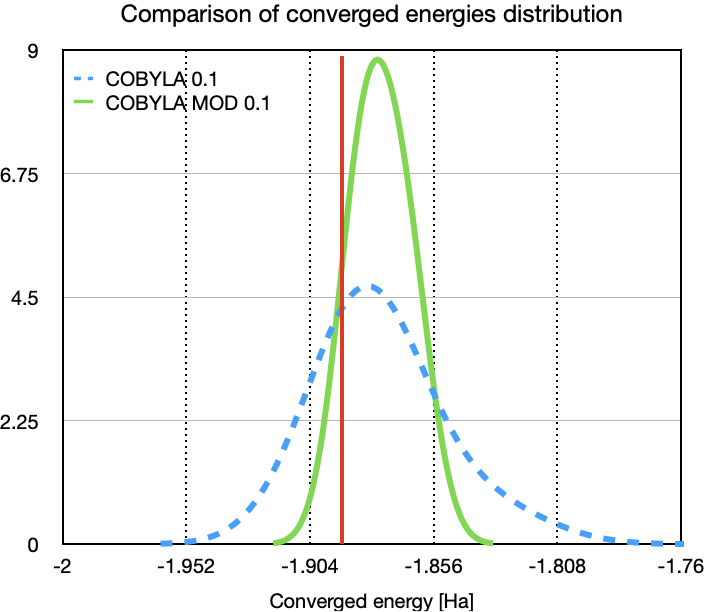}
  \caption{Molecule \ce{H4}. Distribution of the energy of the converged ground state. Comparison of the optimization using COBYLA and COBYLA-MOD. Number of shots is 1000. Default tolerance is 1e-3 Ha.}
  \label{si:fig:opt_h4_comparison}
\end{figure}

In this example we observe a significant improvement in the convergence of the energy using COBYLA-MOD with respect to COBYLA, however they converge to a similar final energy value. Also we observe that for each iteration of the ADAPT-VQE algorithm the number of energy evaluations required for the optimization is lower using COBYLA-MOD. Looking at the error bars associates to the values presented in these figures we also observe that dispersion obtained with COBYLA-MOD is smaller dispersion than ussing COBYLA. This indicates that the results are more reliable using COBYLA-MOD. This  can be observed in figure \ref{si:fig:opt_h4_comparison} where we compare the energy distribution of the final converged energy using both COBYLA and COBYLA-MOD. In this figure we observe that the average energy using COBYLA is slighly smaller but the deviation of the energy distribution is much larger than COBYLA-MOD. 

\begin{figure}[H]
 \includegraphics[width=8cm]{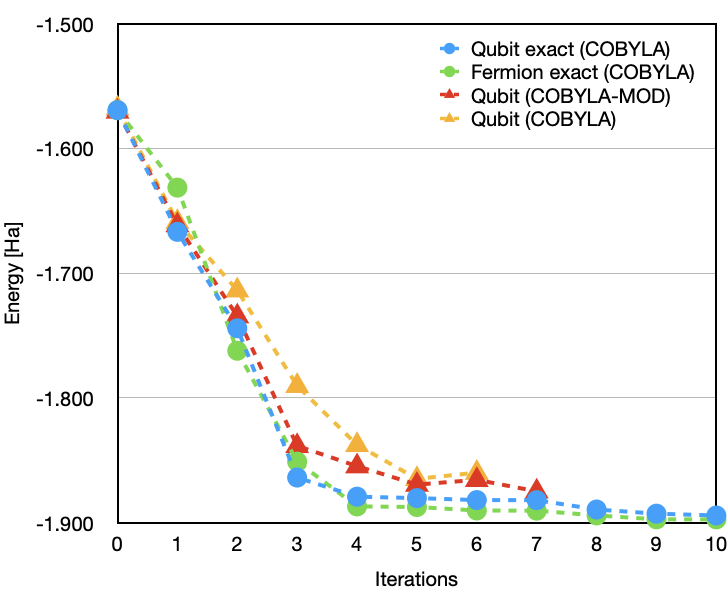}
 \includegraphics[width=8cm]{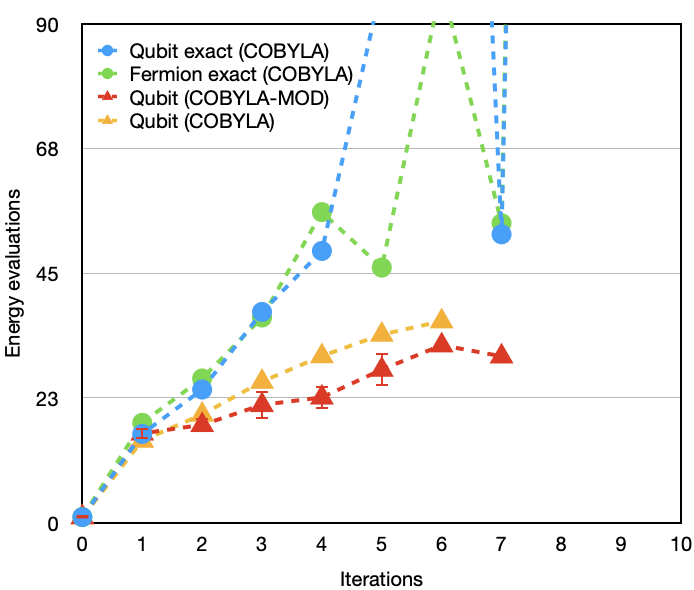}
  \caption{Molecule \ce{H4}. Comparison of the optimization using COBYLA and COBYLA-MOD. Fermion vs qubit operators using a value of 0.1 for rhobeg. Number of shots is 1000. Energy as a function of the iterations (left) and evaluations of the energy (right)}
  \label{si:fig:opt_h4_comparison_ener}
\end{figure}

\begin{figure}[H]
 \includegraphics[width=8cm]{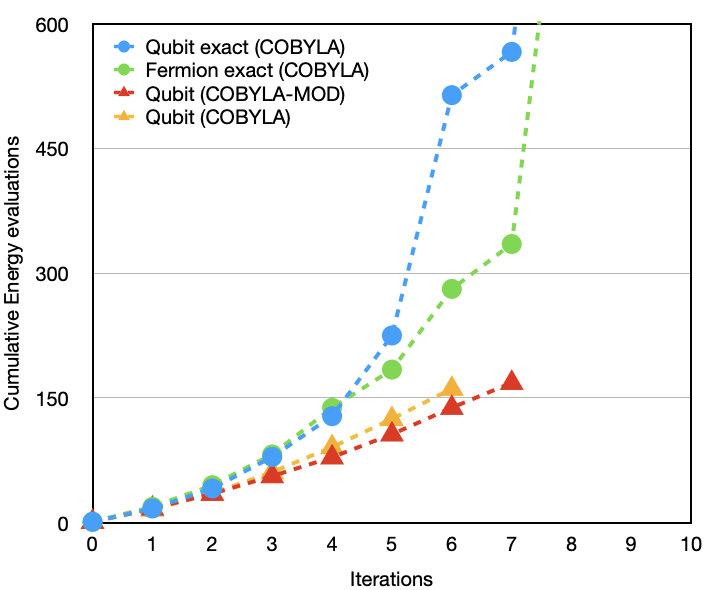}
  \caption{Molecule \ce{H4}. Comparison of the optimization using COBYLA and COBYLA-MOD. Fermion vs qubit operators using a value of 0.1 for rhobeg. Number of shots is 1000}
  \label{si:fig:opt_h4_comparison_cumul}
\end{figure}

\begin{figure}[H]
 \includegraphics[width=8cm]{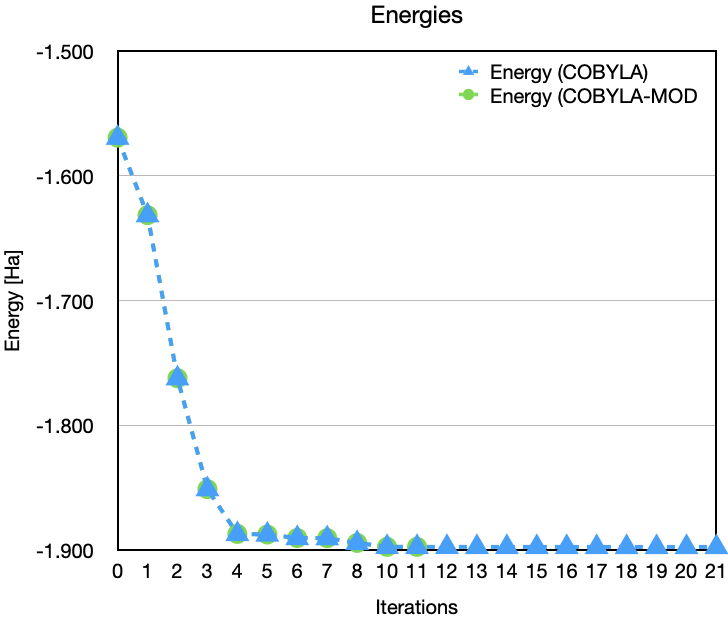}
 \includegraphics[width=8cm]{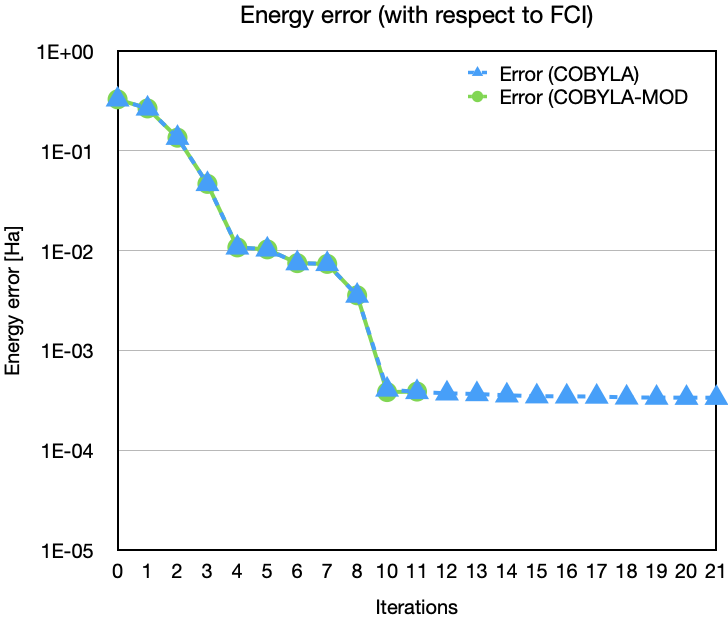}
  \caption{Molecule \ce{H4}. Comparison of the optimization using COBYLA and COBYLA-MOD.  Exact energy evaluation vs ADAPT iterations (left) and energy error with respect to FCI (right)}
  \label{si:fig:opt_h4_comparison_error}
\end{figure}

\begin{figure}[H]
 \includegraphics[width=8cm]{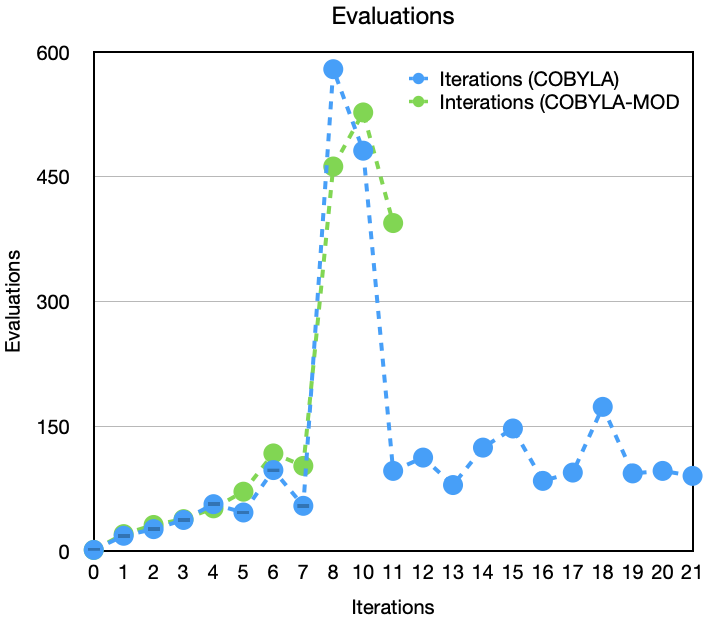}
  \caption{Molecule \ce{H4}. Comparison of the optimization using COBYLA and COBYLA-MOD.  Energy evaluations vs ADAPT iterations}
  \label{si:fig:opt_h4_comparison_iter}
\end{figure}

\section{Circuit optimization}

\begin{table}[H]
\centering
\caption{Circuit depth of the ansatz used at each iteration step of the Qubit-ADAPT using the two different CNOT orientations and the one generated by our algorithm (Optimal). For each version we show also the depth of the transpiled circuit for \texttt{IBM torino} quantum computer (ISA depth).}
\begin{tabular}{cccccccccc}
\toprule
 & \multicolumn{3}{c}{{Standard}} & \multicolumn{3}{c}{{Reverse}} & \multicolumn{3}{c}{{Optimal}} \\
\cmidrule(lr){2-4} \cmidrule(lr){5-7} \cmidrule(lr){8-10}
 \# op. & {depth} & {CNOT} & {ISA} & {depth} & {CNOT} & {ISA} & {depth} & {CNOT} & {ISA} \\
\midrule
0 & 1  & 0  & 1   & 1  & 0  & 1   & 1  & 0  & 1   \\
1 & 10 & 6  & 22  & 8  & 6  & 24  & 8  & 6  & 24  \\
2 & 10 & 12 & 82  & 8  & 12 & 74  & 8  & 12 & 80  \\
3 & 19 & 18 & 126 & 17 & 18 & 132 & 17 & 18 & 132 \\
4 & 35 & 32 & 173 & 33 & 32 & 181 & 33 & 32 & 181 \\
5 & 42 & 38 & 207 & 39 & 38 & 215 & 39 & 38 & 215 \\
6 & 44 & 41 & 197 & 41 & 41 & 215 & 41 & 41 & 215 \\
7 & 57 & 54 & 242 & 54 & 54 & 253 & 54 & 54 & 253 \\
\bottomrule
\end{tabular}
\label{tbl:qubit_depth}
\end{table}

\begin{table}[H]
\centering
\caption{Circuit depth of the ansatz used at each iteration step of the Fermion-ADAPT using the two different CNOT orientations and the one generated by our algorithm (Optimal). For each version we show also the depth of the transpiled circuit for \texttt{IBM torino} quantum computer (ISA depth).}
\begin{tabular}{cccccccccc}
\toprule
 & \multicolumn{3}{c}{{Standard}} & \multicolumn{3}{c}{{Reverse}} & \multicolumn{3}{c}{{Optimal}} \\
\cmidrule(lr){2-4} \cmidrule(lr){5-7} \cmidrule(lr){8-10}
 \# op. & {depth} & {CNOT} & {ISA} & {depth} & {CNOT} & {ISA} & {depth} & {CNOT} & {ISA} \\
\midrule
0 & 1    & 0    & 1    & 1    & 0    & 1    & 1    & 0   & 1    \\
1 & 73   & 48   & 110  & 68   & 48   & 64   & 67   & 46  & 64   \\
2 & 73   & 96   & 232  & 68   & 96   & 149  & 67   & 92  & 149  \\
3 & 696  & 576  & 1244 & 683  & 576  & 1059 & 616  & 514 & 1060 \\
4 & 1110 & 896  & 1949 & 1090 & 896  & 1631 & 979  & 794 & 1628 \\
5 & 1182 & 944  & 2072 & 1157 & 944  & 1719 & 1045 & 840 & 1726 \\
6 & 1254 & 992  & 2074 & 1157 & 992  & 1720 & 1045 & 886 & 1718 \\
7 & 1408 & 1040 & 2183 & 1224 & 1040 & 2293 & 1111 & 932 & 1790 \\
\bottomrule
\end{tabular}
\label{tbl:fermion_depth}
\end{table}

\section{Job Size Limitations in IBM Quantum Computers}\label{sec:circuit_lim_SI}

To estimate the computational limitations of IBM quantum 
computers, we submitted a series of jobs using the
\texttt{Estimator} primitive from the IBM Qiskit Runtime.
These jobs varied in both the number of two-qubit gates and
the number of Pauli observables measured.

The data shown in the plot below reflects the limitations
observed at the time of writing this manuscript. A job was
considered not allowed if the server returned an error at
submission time, indicating that the circuit exceeded current
resource or policy constraints.

\begin{figure}[H]
   \centering
   \includegraphics[width=0.8\textwidth]{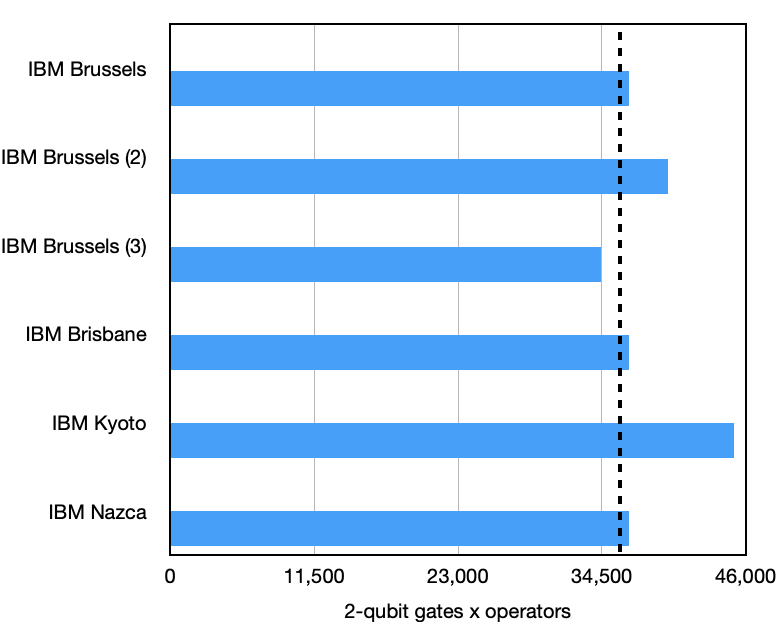}
   \caption{Maximum job size accepted by various IBM quantum computers, measured as the product of the number of two-qubit gates and the number of measured Pauli operators in the observable. IBM Brussels was tested 3 times}
   \label{fig:circuit_limits}
\end{figure}

\section{Quantum hardware experiments}\label{sec:quantum_exp_SI}

\begin{figure}[H]
    \includegraphics[width=6.5cm]{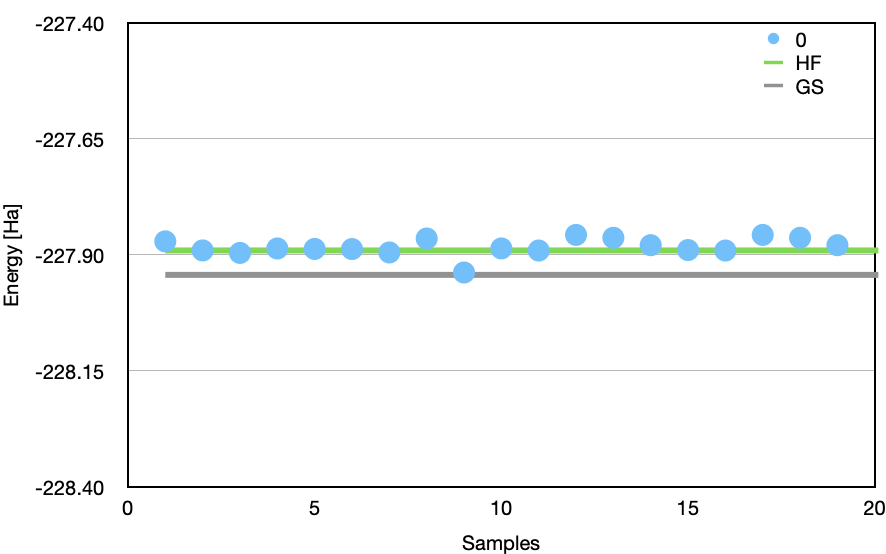}   \includegraphics[width=6.5cm]{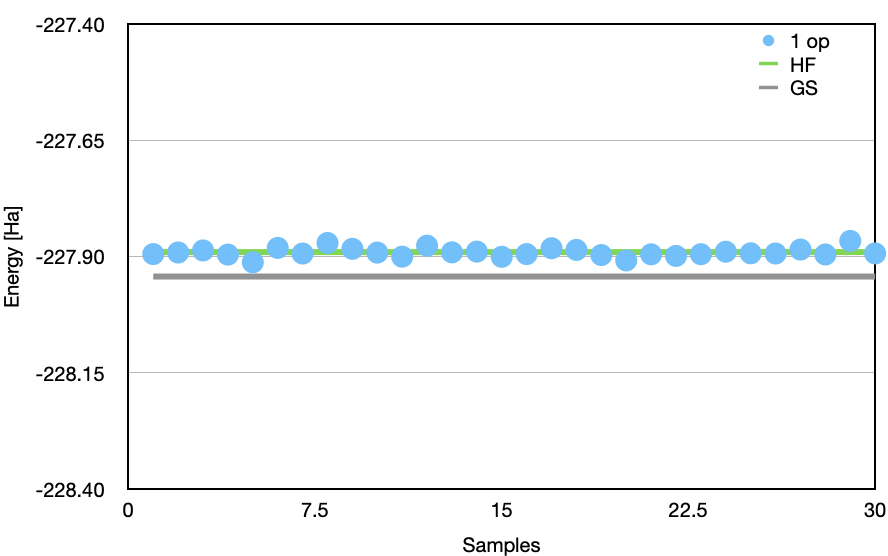}
    \includegraphics[width=6.5cm]{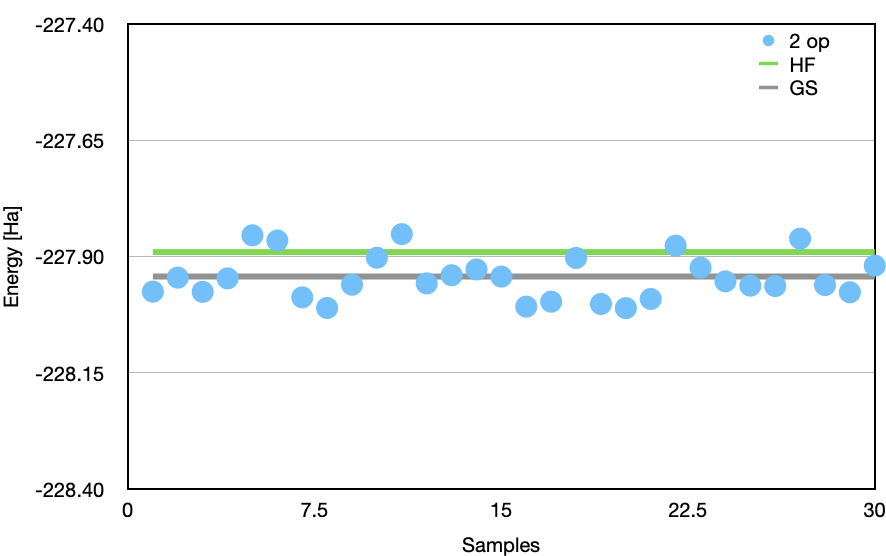}   \includegraphics[width=6.5cm]{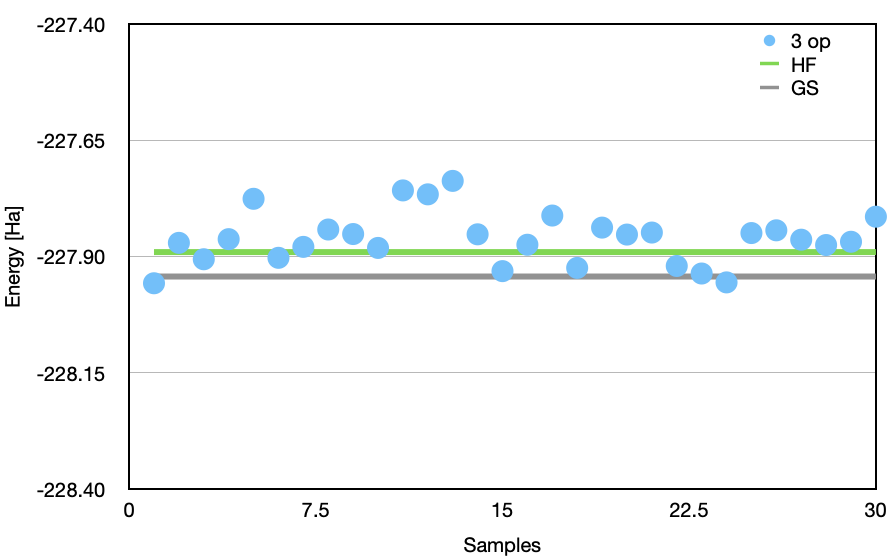} 
    \includegraphics[width=6.5cm]{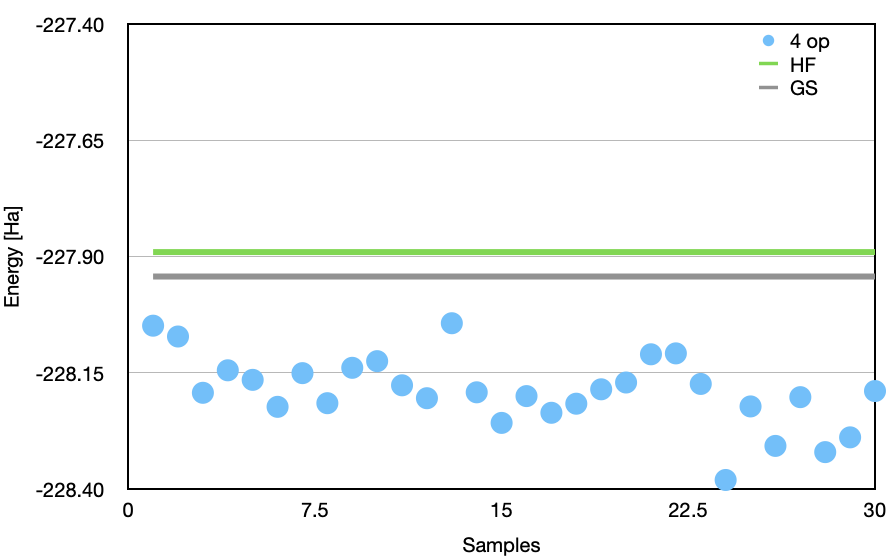}   \includegraphics[width=6.5cm]{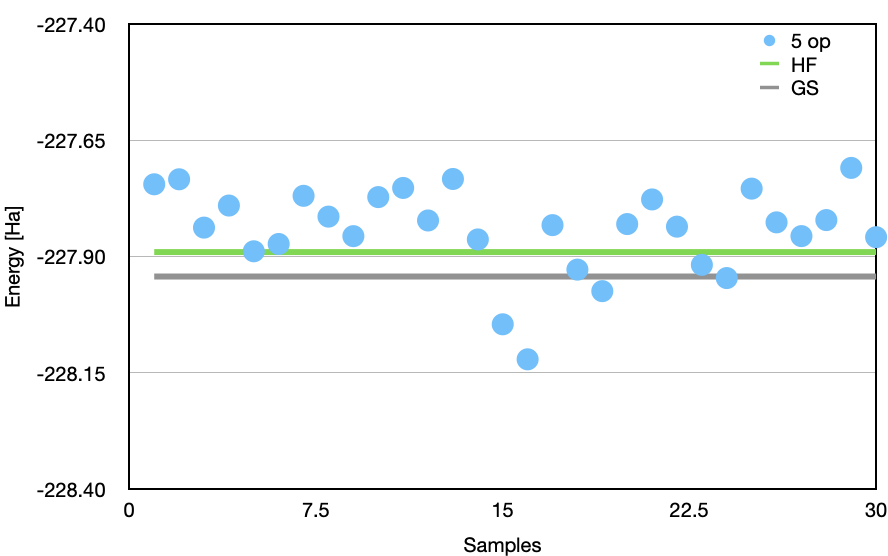} 
    \includegraphics[width=6.5cm]{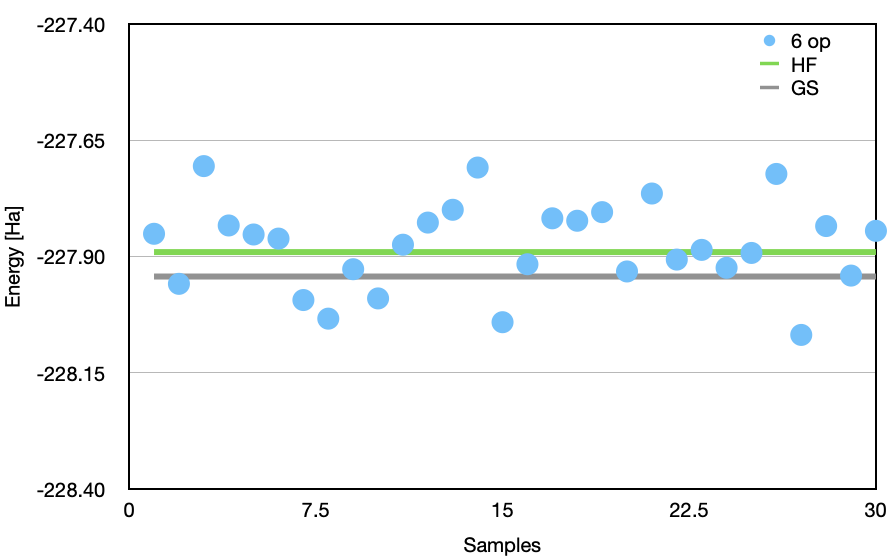}    \includegraphics[width=6.5cm]{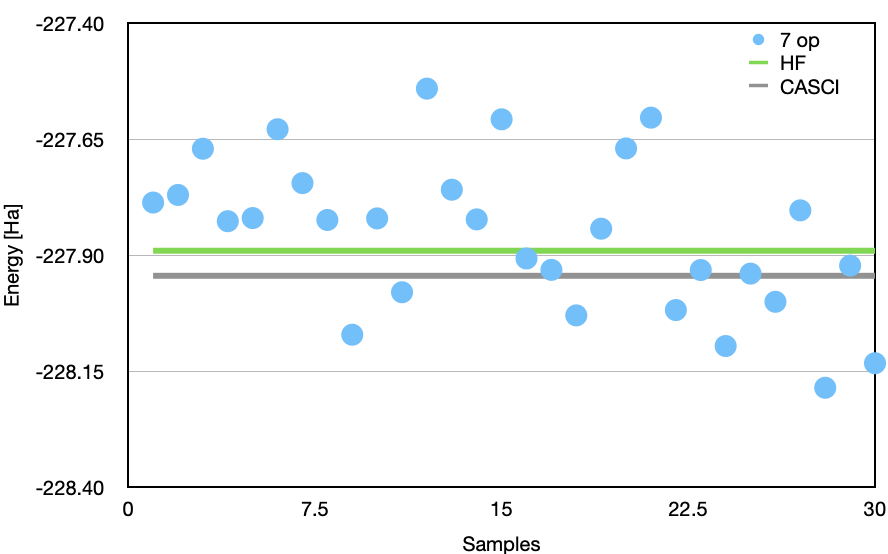} 
  \caption{Energy evaluations (in Ha) with Qubit-ADAPT ans\"atze with 0-7 qubit operators measured in the \texttt{IBM Torino} computer using 9000 shots per sample. The coefficients of the ansatz are obtained from an exact calculation. Horizontal lines indicate the exact HF (green) and FCI with 4 active orbitals and 4 active electrons (gray). 
  The energy difference between HF and FCI (correlation energy) is 53.9 mHa.
  Error bars corresponds to $2\sigma$.}
  \label{fig:samples_rh}
\end{figure}

\section{Instruction execution times}

Instructions times consider in our Qubit-ADAPT simulations with \texttt{Qiskit Aer} simulator are shown in Table~\ref{tbl:instruction_times}, corresponding to:
\begin{itemize}
    \item Parameterized single-qubit rotation gates
    \begin{itemize}
        \item U1: phase shift gate.
        \item U2: single-qubit rotation with two parameters.
        \item U3: general single-qubit unitary rotation with three parameters.
    \end{itemize}
    \item CX: CNOT gate.
    \item Reset: reset operation initializing a qubit back to the $|0\rangle$ state.
    \item Measure: the measurement operation collapses a qubit's state into either $|0\rangle$ or $|1\rangle$ and records the result.
\end{itemize}

\begin{table}[h]
    \centering
    \caption{Instruction execution times (in ns) used in the noise model.}
    \begin{tabular}{lc}
        \hline
        \hline
        {instruction} & {time} \\
        \hline
        U1       & 0        \\ 
        U2       & 50       \\
        U3       & 100      \\
        CX       & 300      \\ 
        Reset    & 1000     \\ 
        Measure  & 1000     \\
        \hline
    \end{tabular}
    \label{tbl:instruction_times}
\end{table}
